\documentclass[fleqn,usenatbib]{mnras}

\usepackage{newtxtext,newtxmath}

\usepackage[T1]{fontenc}

\DeclareRobustCommand{\VAN}[3]{#2}
\let\VANthebibliography\thebibliography
\def\thebibliography{\DeclareRobustCommand{\VAN}[3]{##3}\VANthebibliography}

\usepackage{graphicx}	
\usepackage{amsmath}	
\usepackage{caption}
\usepackage{subcaption}

\title[Disc Evolution in GRO J1655-40 \& LMC X-3]{Accretion Disc Evolution in GRO J1655-40 and LMC X-3 with Relativistic and Non-Relativistic Disc Models}

\author[A. Yilmaz et al.]{
Anastasiya Yilmaz,$^{1,2}$\thanks{E-mail: anastasiya.yilmaz@asu.cas.cz}
Ji\v{r}í Svoboda,$^{1}$
Victoria Grinberg,$^{3}$
Peter G. Boorman,$^{1,4}$
\newauthor
Michal Bursa,$^{1}$
Michal Dov\v{c}iak$^{1}$
\\
$^{1}$Astronomical Institute of the Czech Academy of Sciences, Bo\v{c}n\'\i -II 1401, Praha 4 Prague, 141~00, Czech Republic\\
$^{2}$Astronomical Institute of Charles University, Faculty of Mathematics and Physics, Charles University, V Holešovičkách 2, Prague 8, 180~00, Czech Republic\\
$^{3}$European Space Agency (ESA), European Space Research and Technology Centre (ESTEC), Keplerlaan 1, 2201 AZ Noordwijk, The Netherlands\\
$^{4}$ Cahill Center for Astronomy and Astrophysics, California Institute of Technology, Pasadena, CA 91125, USA
}

\date{Accepted XXX. Received YYY; in original form ZZZ}

\pubyear{2023}

\begin{document}
\label{firstpage}
\pagerange{\pageref{firstpage}--\pageref{lastpage}}
\maketitle

\begin{abstract}
Black hole X-ray binaries are ideal environments to study the accretion phenomena in strong gravitational potentials. These systems undergo dramatic accretion state transitions and analysis of the X-ray spectra is used to probe the properties of the accretion disc and its evolution. In this work, we present a systematic investigation of $\sim$1800 spectra obtained by RXTE PCA observations of GRO J1655-40 and LMC X-3 to explore the nature of the accretion disc via non-relativistic and relativistic disc models describing the thermal emission in black-hole X-ray binaries. We demonstrate that the non-relativistic modelling throughout an outburst with the phenomenological multi-colour disc model \texttt{DISKBB} yields significantly lower and often unphysical inner disc radii and correspondingly higher ($\sim$50-60\%) disc temperatures compared to its relativistic counterparts \texttt{KYNBB} and \texttt{KERRBB}. We obtained the dimensionless spin parameters of $a_{*}=0.774 \pm 0.069 $ and $a_{*}=0.752 \pm 0.061 $ for GRO J1655-40 with \texttt{KERRBB} and \texttt{KYNBB}, respectively. We report a spin value of $a_{*}=0.098 \pm 0.063$ for LMC X-3 using the updated black hole mass of 6.98 ${M_{\odot}}$. Both measurements are consistent with the previous studies. Using our results, we highlight the importance of self-consistent modelling of the thermal emission, especially when estimating the spin with the continuum-fitting method which assumes the disc terminates at the innermost stable circular orbit at all times.
\end{abstract}




\section{Introduction}
Black Hole X-ray Binaries (BHXRBs) are a sub-class of X-ray binaries with a black hole as the central compact object and most of them are observed to be transient sources. The spectral properties of these accreting systems are primarily described depending on the relative contribution of two distinct emission components: thermal and non-thermal. The thermal component of the X-ray spectrum is composed of radiation at different radii in the accretion disc \citep{Shakura73} and the non-thermal contribution to the spectrum arises due to the thermal Comptonisation of the "seed" photons from the innermost regions of the accretion disc by a hot corona (an optically thin cloud of electrons close to the black hole) usually responsible for the emission above $\sim $ 5 keV \citep{Sunyaev80, Zdziarski85, Chakrabarti95, Narayn96, Zdziarski04}. Superimposed on the continuum, a third component, the relativistic reflection appears in the event of non-thermal emission being reprocessed by the optically thick disc.

While only a few systems are observed to be persistently active, a big majority of the BHXRBs spend their time in a quiescent state where the total luminosity in all bands is very low and their evolution as a system is mainly described by properties of accretion states. To date, several accretion states were identified in BHXRBs (see, e.g., \citet{Fender04} and \citet{Remillard06}) that rapidly change over time depending on the relative dominance of the thermal and non-thermal components of the spectral continuum. Once the outburst begins, the system starts to transit across three predominantly identified distinct accretion states according to the X-ray intensity and X-ray spectral hardness parameters. These accretion states are usually tracked using a hardness-intensity diagram (HID) and many BHXRBs are found to follow a clear "q-shape" \citep{Fender04, Dunn10}. For a typical outburst, these accretion states are identified as the low/hard state (LHS), high/soft state (HSS) and the steep powerlaw or intermediate state. 

At the very beginning and the end of the outburst, the binary system is observed to be in the LHS state where the luminosity is significantly lower and most of the X-ray flux is dominated by the hard emission from an X-ray corona which can be described by a powerlaw with a relatively flat photon index with $\Gamma \sim 1.5-2$ and a weak disc component characterized by lower disc temperatures with $k T<1$ keV \citep{Poutanen98}. The increase in the X-ray luminosity throughout the rise of the outburst is usually followed by a rise in the radio luminosity. This "compact" radio emission is associated with synchrotron radiation from a steady relativistic jet with a flat or in some cases inverted radio spectrum (see \citet{Blandford79, Fender01, Fender14}). Independent of the increasing luminosity by a few orders of magnitude (both in radio and X-rays), the spectrum remains hard throughout the LHS \citep{Corbel00,Corbel03}. As the accretion rate increases, the system then transitions to the steep powerlaw or the intermediate state where the luminosity is high and the spectrum is dominated by a steep powerlaw ($\Gamma \sim 1.9$) \citep{Remillard06, Belloni16}. This state can often be divided into two arms with the hard intermediate state (HIMS) and the soft intermediate state (SIMS) where the source transitions through a powerlaw-dominated state to a more thermal state before transitioning to the HSS. During this transition to the HSS, the system crosses a "jet line" which is described as a point throughout the outburst where the compact radio jet is quenched \citep{Fender99, Corbel00, Russel11}. During the HSS, where the X-ray luminosity can increase by orders of magnitude and the big majority of the flux is observed to be emitted from the accretion disc with a spectrum that is primarily characterized by a multi-temperature blackbody with a peak temperature of $kT \sim $ 1-1.5 keV with an additional powerlaw with a photon index of $\Gamma \sim 2.5$ or even softer. In contrast to the substantial increase in the X-ray luminosity during the rise to the HSS, the radio emission starts to decrease and eventually vanishes as the radio jet is quenched in this state. The softening of the spectrum from the LHS to HSS occurs relatively faster compared to the time spent in the LHS \citep{Dunn10}. The third state, the steep powerlaw or the intermediate state, is entered as the powerlaw also starts to dominate over the thermal disc emission with $\Gamma \sim 2.5-3$ but the spectrum continues to be very soft. As the dominance of the disc continues to fade away, the system returns to the LHS completing the q-path at the end of the outburst before re-entering the quiescent state. Such sharp increases in the X-ray luminosity throughout the outburst while transitioning to the HSS are primarily explained by rapidly increasing accretion rates, $\dot{M}$, in a geometrically thin, optically thick disc originally characterized by \citet{Shakura73}.

The most extensive global spectral study of BHXRBs was introduced by \citet{Dunn10} where they analysed a sample of 25 known low-mass BHXRBs observed by the RXTE mission with multiple outbursts covered adopting a non-relativistic multi-colour disc model with an additional powerlaw component. They extended their work in \citet{Dunn11} focusing on the study of the long-term evolution of accretion discs in BHXRBs throughout the outburst phases of the same sample. For the majority of BHXRBs in their sample, the results followed the expected theoretical $L_{\rm {Disc }} \propto {T}^{4}$ relation with some exceptions at the low and high ends of ${L_{\rm {Disc }}}$ with arms extending out to significantly higher temperatures compared to observations with similar ${L_{\rm {Disc }}}$ values. It was argued that a changing colour correction factor ${f_{\rm {col }}}$ could account for such deviations from the theoretically expected trend but they didn't suspect it to be such a huge influence on the measurements of inner disc radius and temperature. This colour correction factor accounts for spectral hardening of the accretion disc emission which can deviate from a "perfect" blackbody emission due to reprocessing of the disc emission in the disc atmosphere and relates the colour temperature to the effective temperature as $f_{\mathrm{col}} = T_{\mathrm{col}} / T_{\mathrm{eff}}$. For a typical \citep{Shakura73} disc with sufficiently high temperature at the disc surface, electron scatterings start to dominate over any absorption leading to photons emitted from the disc to be up-scattered to higher temperatures. These up-scattering processes produce an emissivity profile modified by the colour correction factor. This colour correction factor can be considered to vary with the accretion state, though constrained measurements of ${f_{\rm {col }}}$ become very difficult from X-ray spectra (for a more in-depth review, see \citet{Salvesen21} and references therein). Similar results for individual sources were previously reported by \citet{Gier04}, \citet{Done07} and \citet{Dunn08}. They showed that a similar trend was present in ${R_{\rm {in}}}$ - $T$ where they reported notably low ${R_{\rm {in}}}$ values at the beginnings and ends of the outbursts in response to their ${L_{\rm {Disc}}}$ - $T$ findings. 

In the most widely accepted picture (see the reviews by \citet{Fender04, Homan05, Remillard06, Done07, Dunn10, Belloni11, Dunn11, Fender12, Fender16}), the radius and the temperature of the innermost region of the accretion disc change throughout the outburst. While transitioning from the beginning of the outburst to the HSS with increasing luminosity, the innermost region of the disc starts to get closer to the black hole \citep{Gier04, Steiner10, Reynolds13}, eventually resulting in a significant increase in the disc temperature peaking at $\mathrm{k T} \sim 1-1.5\: \mathrm{keV}$. During the peak of the HSS, the disc can extend all the way down to the innermost stable circular orbit (ISCO) with a contrastingly cold disc truncated at much larger radii in the LHS at the beginning and end of an outburst \citep{Narayan95, Esin97, D'Angelo08, Reis10, Plant15}. The exact accretion disc geometry throughout the LHS remains less understood compared to HSS. Both in relativistic and non-relativistic solutions, a "zero-torque" boundary condition at the ISCO is assumed \citep{Shakura73, Novikov73} while, in reality, magnetic fields strong enough to exert such torques might exist but can not be strong enough to dominate over the mass accretion \citep{Krolik99, Gammie99}. When magnetic torques at the inner boundary are taken into account, the radiative efficiency of the accretion disc increases significantly.

This changing inner disc radius was challenged by XMM-Newton and RXTE observations of SWIFT J1753.5-0127 where \citet{Miller06} reported an inner disc radius $R_{\text {in }} \leq 6 G M / c^{2}$ in the LHS for all of their spectral models. A similar trend was later reported by \citet{Reis10} for 8 black hole X-ray binary systems in the LHS while the disc temperatures remained below 0.3 keV.

The radius of the ISCO is determined primarily by the spin and mass of the black hole, 
\begin{equation}\label{r_isco}
	R_{\text {ISCO}}=r_{\mathrm{g}}\left\{3+A_{2} \pm\left[\left(3-A_{1}\right)\left(3+A_{1}+2 A_{2}\right)\right]^{1 / 2}\right\}
\end{equation}
\begin{equation}\label{A_1}
	A_{1}=1+\left(1-a_{*}^{2}\right)^{1 / 3}\left[\left(1+a_{*}\right)^{1 / 3}+\left(1-a_{*}\right)^{1 / 3}\right]
\end{equation}
\begin{equation}\label{A_2}
	A_{2}= \left(3 a_{*}^{2}+A_{1}^{2}\right)^{1 / 2} 
\end{equation}

\begin{equation}\label{A_2}
	a_{*}=\frac{|\mathbf{J}| c}{G M^{2}} 
\end{equation}

where $J$ is the black hole angular momentum, $G$ is the gravitational constant, $c$ is the speed of light, $M$ is the black hole mass, $r_{\rm g}= G M / c^{2}$ is the gravitational radius and the upper and lower signs correspond to a prograde disc ($a_{*}>0$) and a retrograde disc ($a_{*}<0$), respectively. By this relation, an accretion disc around a maximally spinning black hole with $a_{*}=1$ can extend all the way down to just $R_{\text {ISCO }}\left(a_{*}=1\right)= r_{\rm g}$ while in the canonical Schwarzschild case with $a_{*}=0$ it can be only as close as $R_{\text {ISCO }}\left(a_{*}=0\right)=6 r_{\rm g}$.

In this paper, we focus on several outbursts of GRO J1655-40 and LMC X-3 and perform a detailed spectral analysis of all of the publicly available RXTE observations spanning nearly 16 years of observing time. The structure of this paper is as follows: in Section~\ref{sources}, we summarize the properties of GRO J1655-40 and LMC X-3, outline the selected observations and data reduction process, provide a review of the spectral modelling of accretion discs in black hole X-ray binaries and summarize the spectral analysis using these models. In Section~\ref{results}, we present the results of our analysis with each model and provide a discussion in Section~\ref{discussion}. Section~\ref{summary} provides a summary of our conclusions based on our analysis. 

\section{Sources, Data Reduction and Analysis}\label{sources}

\subsection{GRO J1655-40}
GRO J1655-40, one of the most extensively studied Galactic low mass black hole X-ray binaries (LMBXRBs)\citep{Hjellming95, Orosz+97, Abramowicz01}, was discovered in 1994 by the Burst and Transient Source Experiment (BATSE) onboard the Compton Gamma Ray Observatory \citep{Zhang94}. \citet{Orosz97} measured a black hole with a mass of ${ M_{\mathrm{BH}}}=6.3 \pm 0.5 \: {M_{\odot}}$ and a companion star of ${ M_{*}}=2.4 \pm 0.4 \: { M_{\odot}}$ based on optical observations with an orbital period of 2.6 d \citep{Greene01}. Based on optical/NIR observations, \citet{Beer02} reported ${M_{\mathrm{BH}}}=5.4 \pm 0.3 \: {M_{\odot}}$ and ${M_{*}}=1.45 \pm 0.35 \: {M_{\odot}}$. The inclination angle has been reported as $i=70.2^{\circ} \pm 1.9^{\circ}$ \citep{Kuulkers00} while radio observations revealed a difference of $15^{\circ}$ between the binary orbital plane inclination angle and the jet inclination angle \citep{Orosz97, Maccarone02}. Although a distance of $\sim$ $3.2 \pm 0.2$ kpc is commonly used, \citet{Foellmi06} argued that it should be lower than 1.7 kpc. In this study, we adopt d $\sim$ $3.2 \pm 0.2$ kpc based on \citet{Hjellming95}. Different methods to measure the spin on GRO J1655-40 revealed a controversy, with $0.65< a_{*} <0.75$ obtained from spectral continuum fitting  with ASCA and RXTE observations \citep{Shafee06} while Fe emission line profile predicts much higher values, with a lower limit on the spin at $a_{*}=0.9$ \citep{Reis09} based on XMM-Newton observations. On the contrary to these measurements pointing towards a high spin case for GRO J1655-40, timing analysis of the entire set of RXTE data using the relativistic precession model \citep{Stella98, Stella99} resulted in a much smaller spin parameter $a_{*} = 0.290 \pm 0.003$ \citep{Motta14}. During the 2005 outburst, GRO J1655-40 was observed mainly in HSS except for unusually soft or hyper-soft states \citep{Uttley15} where detection of a strong accretion disc wind with a line-rich X-ray absorption spectrum was reported using Chandra X-ray HETG \citep{Miller08}. This hyper-soft state is suggested to arise from a Compton-thick accretion disc wind and might even indicate intrinsic luminosities above the Eddington limit due to the obscuration \citep{Neilsen16}. While the underlying mechanism is still not well understood, one of the explanations to describe the nature of this strong disc wind is based on a hybrid magnetically/thermally driven mechanism \citep{Neilsen12}.

\subsection{LMC X-3} 
LMC X-3, a bright and persistent low mass X-ray binary system in the Large Magellanic Cloud (LMC), was discovered by UHURU in 1971 \citep{Leong71} located at a distance of d $\sim$48.1 kpc \citep{Orosz09}. The system harbours a black hole with an estimated mass of ${ M_{\mathrm{BH}}}=6.98 \pm 0.56 { M_{\odot}}$ and a companion star categorized as a B3 V star with an estimated mass of ${M_{*}}=3.63 \pm 0.57 {M_{\odot}}$. This highly variable binary system has an inclination angle $i=69^{\circ}\pm 0.72^{\circ}$ \citep{Orosz14} and has been observed to be dominantly in the HSS \citep{Nowak01} except for a few occasions with observations in LHS \citep{Boyd00, Wilms01} with a few anomalous low/hard state (ALS) observations with a dramatic drop in the total X-ray flux down to $\sim 1 \times 10^{35}$ erg ${\rm \mathrm{s}}^{-1}$ in 2-10 keV range \citep{Smale12,Torpin17} making it an object of interest to study the accretion phenomena and various accretion disc models in BHXRBs \citep{Davis06, Kubota10, Straub11}. LMC X-3 stands out as a unique object due to the almost always-on and dominantly soft nature of its X-ray observations. Its nearly constant inner disc radius \citep{Steiner10} puts LMC X-3 in the spotlight for spin measurements using X-ray spectral continuum fitting. Spin measurements to date using the X-ray spectral continuum fitting method with models based on a thin disc structure revealed a relatively low spin parameter in the range of $0.22<a_{*}<0.41$ \citep{Davis06, Steiner14, Bhuvana21, Bhuvana22} while spectral modelling adopting a relativistic slim disc model with different viscosity and colour correction parameters gave much higher spin values of $a_{*} \sim 0.7$ \citep{Straub11}. We provide a summary of all of the physical parameters adopted in this paper for each source in Table~\ref{table:src}.

\begin{table}
\centering
	\caption{Parameter Values for LMC X-3 and GRO J1655-40} 
	\label{table:src}
	\begin{tabular}{p{0.09\textwidth}p{0.08\textwidth}p{0.07\textwidth}p{0.08\textwidth}p{0.06\textwidth}} 
		\hline \hline Source & $M_{\mathrm{BH}}$ & $D$  & $i$  &$N_{\mathrm{H}}$  \\
		& $(M_{\odot})$ & $(\mathrm{kpc})$& $(\mathrm{deg})$ & $(10^{22} \mathrm{~cm}^{-2})$\\
        \hline 
        LMC X-3 & $6.98 \pm 0.56 ^{{a}}$ & 48.1 $^{{b}}$ & $69.0\pm 0.72 ^{{a}}$ & $0.04^{{c}}$ \\
        GRO 1655-40 & $6.3 \pm 0.5 ^{d}$ & $3.2 \pm 0.2^{e}$ & $70.2 \pm 1.9 ^{f}$  & $0.8^{d}$\\
	\hline \hline
	\end{tabular}
    $^\mathrm{a}$ \citet{Orosz14}, $^\mathrm{b}$ \citet{Orosz09}, $^\mathrm{c}$ \citet{Page03}, $^\mathrm{d}$ \citet{Orosz97}, $^\mathrm{e}$ \citet{Hjellming95}, $^\mathrm{f}$ \citet{Kuulkers00}
\end{table}

\subsection{Observations}
The Rossi X-Ray Timing Explorer (RXTE) observed LMC X-3 and GRO J1655-40 extensively throughout the entire mission spanning multiple outbursts between 1997-2011 and 1996-2012, respectively. For the spectral fitting, we used a total of more than 1800 observations publicly available on HEASARC (High Energy Astrophysics Science Archive Research Center) archive
. The full list of proposal numbers of the observations used can be found in Table~\ref{table:obs}.

\begin{table}
\centering
	\caption{ Proposal numbers and observation dates of RXTE observation used in spectral analysis (in ascending order).} 
	\label{table:obs}
	\begin{tabular}{p{0.09\textwidth}p{0.099\textwidth}p{0.09\textwidth}p{0.09\textwidth}}
		\hline \hline
		 Proposal Number & Start Date & End date & Number of Observations \\
		\hline
		&\textbf{GRO J1655-40}\\
		\hline
		10255 & 1996-05-09 & 1997-01-26 & 31 \\
		10261 & 1996-05-14 & 1996-06-30 & 1\\
		10263 & 1996-05-12 & 1996-05-12 & 1\\
		10404 & 1996-03-14 & 1996-03-14 & 15\\
		20187 & 1996-11-07 & 1996-11-07 & 1\\
		20402 & 1997-02-26 & 1997-09-11 & 32\\
		90019 & 2005-03-13 & 2005-03-13 & 1\\
		90058 & 2005-02-20 & 2005-03-02 & 6\\
		90428 & 2005-02-25 & 2005-03-03 & 11\\
		90704 & 2005-03-10 & 2005-03-10 & 2\\
		91404 & 2005-03-04 & 2005-03-07 & 6\\
		91702 & 2005-03-07 & 2005-11-11 & 473\\ 
		91704 & 2005-09-24 & 2005-09-24 & 3\\
		& &  Total: & 583\\
        
            \hline
		&\textbf{LMC X-3}\\
		\hline
		10252 & 1996-02-09 & 1996-10-24 & 67 \\
		20188 & 1996-11-30 & 1997-12-12 & 27\\
		20189 & 1997-08-29 & 1997-09-02 & 2\\
		30087 & 1998-01-06 & 1998-09-30 & 17\\
		40066 & 1998-12-08 & 1998-12-14 & 19\\
		40067 & 1999-01-19 & 1999-09-19 & 26\\
		50411 & 2000-05-05 & 2000-10-22 & 12\\
		60097 & 2001-03-09 & 2002-02-27 & 57\\
		60098 & 2001-06-30 & 2001-07-07 & 4\\
		80103 & 2003-10-04 & 2004-03-10 & 73\\
		80118 & 2004-01-07 & 2004-01-12 & 24\\
		90099 & 2004-04-20 & 2004-05-08 & 12\\
		91105 & 2005-03-07 & 2007-03-16 & 189\\
		92095 & 2006-03-04 & 2007-06-27 & 255\\
		93113 & 2007-06-29 & 2008-12-23 & 156\\
		93114 & 2007-06-28 & 2007-07-09 & 11\\
		94113 & 2008-12-27 & 2009-12-29 & 102\\
		95113 & 2010-01-03 & 2010-12-29 & 103\\
		96113 & 2011-01-02 & 2011-12-28 & 98\\
		& &  Total: & 1254\\
		\hline \hline
		
	\end{tabular}
\end{table}
\subsection{RXTE Data Reduction}

We extracted the source spectra in the \texttt{standard2f} mode which provides the optimal spectral resolution and generated response matrices and background spectra following the standard procedure in the RXTE cookbook\footnote{\url{https://heasarc.gsfc.nasa.gov/docs/xte/recipes/cook_book.html}} using \texttt{FTOOLS/HEASOFT} 
version 6.29\footnote{\url{https://heasarc.gsfc.nasa.gov/docs/software/lheasoft/}} 
software package. We also removed data lying within 10 minutes of the South Atlantic Anomaly. In order to avoid any variations during the data reduction process, we only used the Proportional Counter Unit-2 (PCU-2) of  RXTE’s  Proportional  Counter Array (PCA, \citet{Jahoda06}) with all of its layers included since it was almost always operating throughout the RXTE mission and has the best calibration among other PCUs on-board PCA. We then carried out the  calibration of data using publicly available tool \texttt{pcacorr} \citep{Garcia14} and applied an additional 0.1\% systematic errors to account for uncertainties in the telescope’s response following \citet{Garcia14}. 

\subsection{Spectral Modelling of the Accretion Discs in Black Hole X-ray Binaries}\label{modelling}
For a non-relativistic \citet{Shakura73} accretion disc, the spectrum is assumed to be a sum of blackbody spectra as measured at specific disc radii, $R$, with a characteristic temperature $T_{\mathrm{eff}}(R)$,
\begin{equation}\label{temp}
	T_{\mathrm{eff}}(R)=\left\{\frac{3 G M \dot{M}}{8 \pi R^{3} \sigma}\left[1-\left(\frac{R_{\mathrm{in}}}{R}\right)^{1 / 2}\right]\right\}^{1 / 4}
\end{equation}

where $\dot{M}$ is the mass accretion rate, $M$ is the mass of the black hole, $\sigma$ is the Stefan-Boltzmann constant, $G$ is the gravitational constant and $R_{\mathrm{in}}$ is the innermost radius of the accretion disc. The spectrum resulting from this sum is then a multi-colour disc blackbody, with a peak temperature $T_{\mathrm{eff}, \max }$ assumed to be radiated away from the region of the accretion disc closest to the black hole. 

One of the most widely used non-relativistic accretion disc models implemented in \texttt{XSPEC} is \texttt{DISKBB}  \citep{Mitsuda84, Makishima86}, where the model describes a non-relativistic multi-temperature \citet{Shakura73} accretion disc solution with a finite torque at the inner edge of the disc $R_{\mathrm{in}}$ \citep{Gierlinski99}. The model has two free parameters, a characteristic disc temperature and a normalisation factor which is 
described as
\begin{equation}\label{normalization-diskbb}
\mathcal{N}=\left[\frac{R_{\mathrm{in}}}{D_{10}}\right]^{2} \cos \theta
\end{equation}

where $D_{10}$ is the distance to the source in units of 10 kpc and $\theta$ is the inclination angle of the disc. Assuming perfect blackbody radiation arising from specific radii of the accretion disc, the observed total flux is calculated as 

\begin{equation}\label{f_tot}
F_{\mathrm{Tot}}=\mathcal{N} \frac{4 \pi E^{3}}{h^{3} c^{2}} \int_{1}^{\infty} \frac{r}{e^{E/kT_{\mathrm{eff}}(r)}-1}\mathrm{~d} r
\end{equation}

where $E$ is the energy of the photon, $h$ is the Planck's constant, $k$ is the Boltzmann constant, $c$ is the speed of light and $r=\frac{R}{R_{\mathrm{in}}}$.

Any attempt to develop a better understanding of the inner structure of the disc and the nature of the accretion flow should require a more sophisticated spectral analysis which includes a detailed treatment of relativistic effects acting on the accretion disc in the close vicinity of a black hole. For that purpose, there have been many attempts to develop better-suited spectral models. In the default \texttt{XSPEC} environment, \texttt{KERRBB} \citep{Li05} stood out as a comprehensive spectral model to date. It uses the ray-tracing method to calculate the observed spectrum of a geometrically thin optically thick Keplerian disc around a Kerr black hole while accounting for the relativistic processes taking place in the disc, i.e. light bending, self-irradiation or the returning radiation, gravitational redshift, frame dragging and Doppler boost. This ray-tracing technique is applied to calculate the spectrum assuming that the innermost edge of the disc is always at $R_{\text {ISCO }}$. \texttt{DISKBB} measures the maximum disc temperature from the innermost radius of the accretion disc assuming it is a perfect blackbody emission. This assumption is challenged when the complex electron scatterings in the accretion disc atmosphere are considered. As these processes start to dominate over the free-free absorption, the colour temperature rises and becomes larger than the effective temperature of the accretion disc. By default, \texttt{KERRBB} has $f_{\mathrm{col}} = T_{\mathrm{col}} / T_{\mathrm{eff}}$ as an additional parameter. The most widely implemented value stands at $f_{\mathrm{col}} = 1.7$ and we adopt this value throughout the analysis unless stated otherwise.

\texttt{KYNBB} \citep{Dovciak08} was developed as an extension to the original non-axisymmetric general relativistic spectral model package \texttt{KYN}\footnote{\url{https://projects.asu.cas.cz/stronggravity/kyn}} by \citet{Dovciak04} as an external model to \texttt{XSPEC} environment describing an accretion disc around a black hole with spin values spanning the whole parameter space from -1 to 1. Similar to the case of \texttt{KERRBB}, \texttt{KYNBB} makes use of the ray-tracing method for photon paths in vacuum in a Kerr space-time to compute the transfer functions for the tables used in the model, describing a \citet{Novikov73} accretion disc. In contrast to \texttt{KERRBB}, \texttt{KYNBB} offers the option to define the inner disc radius at a different distance than ISCO. One also has the freedom of including polarization calculations \citep{Mikusincova23,Taverna22}, calculating the spectrum only for a certain part of the accretion disc and defining a range on the disc radius from which the spectrum will be calculated. For the purpose of our analysis, polarization calculations are switched off (Stokes parameter set at 0) and the model calculates the spectrum the same way as \texttt{KERRBB} with the only difference of the innermost edge of the accretion disc no longer terminating at $R_{\rm ISCO}$ (with ms switch set at 1). See \citet{Yilmaz23} for a more detailed comparison of \texttt{KERRBB} and \texttt{KYNBB}. We do not apply any smoothing on the spectrum (smooth parameter set at 0) and assume no obscuring cloud by setting r$_{\rm cloud}$=0. Similar to \texttt{DISKBB}, the normalisation parameter of \texttt{KYNBB} is defined as
\begin{equation}\label{normalization-kynbb}
\mathcal{N}=\frac{1}{D_{10}^{2}}
\end{equation}
where $D_{10}$ is the distance to the source in units of 10 kpc. The list of physical parameters and values used for our analysis can be found in Table~\ref{table:model_pars}.

\subsection{Spectral Analysis}\label{spectral_analysis}

For spectral analysis, we used the \texttt{HEASOFT} (v.6.29) package \texttt{XSPEC (v.12.12)} \citep{Arnaud96} and \texttt{PyXspec}\footnote{\url{https://heasarc.gsfc.nasa.gov/docs/xanadu/xspec/python/html/}}, a Python interface to the \texttt{XSPEC}. RXTE PCA covers the energy range of 2-60 keV while the count statistics become significantly poorer beyond the range of 3-25 keV. We, therefore, restricted the energy range to 3-25 keV for all of the PCA spectra. We used \texttt{grppha} to group the spectra to have 25 counts per bin. All of the models used for the spectral analysis consist of a thermal disc component, either relativistic or non-relativistic, and a simple powerlaw component accounting for the non-thermal component of the continuum (\texttt{POWERLAW} in \texttt{XSPEC}). We use \texttt{TBABS} to account for the ISM absorption. For convenience, each model (\texttt{TBABS$\times$(DISKBB+POWERLAW)}, \texttt{TBABS$\times$(KERRBB+POWERLAW))} and \texttt{TBABS$\times$(KYNBB+POWERLAW)} will be referred to as the specific model used for the disc component. We adopted a selection criteria by assessing reduced $\chi^{2}$ ($\chi^{2}$/d.o.f.) values and applied a cut at reduced $\chi^{2}=2.0$ excluding the observations that lie outside this range. We used the normalisation parameter of \texttt{DISKBB}, $\mathcal{N}$, (see  equation~(\ref{normalization-diskbb})), to calculate the inner disc radius and apply no additional corrections to any parameter value obtained from the fitting. All of the errors on model parameters presented throughout the paper are calculated at the 90\% confidence level.

Throughout our analysis with \texttt{KYNBB}, we assumed an emission integrated above the measured $R_{\mathrm{in}}$ covering the entire disc with an exponential radial grid and defined the outer edge of the non-zero disc emissivity as 1000 $R_{\mathrm{g}}$. We did not apply any smoothing on the calculated spectrum. We set the number of radial grid points on the disc at the default value of 150 for sufficient resolution of the spectrum and switched off the parameters of the model regarding an obscuring cloud and polarization calculations. We summarize the values of relevant parameters as adopted in the spectral analysis in Table~\ref{table:model_pars}.

We used disc fractions to distinguish observations based on how much the disc component is dominating over powerlaw following \citet{Kording06, Dunn08, Dunn10} with the disc fraction defined as 
\begin{equation}\label{disk_fraction}
 {F_{\rm Disc}}=\frac{L_{\rm Disc} \rm{(0.001-100.0~keV)}}{L_{\rm Disc} \mathrm{(0.001-100.0~keV)}+L_{\rm Power} \mathrm{(1.0-100.0~keV)}}
\end{equation}
 
where $L_{\mathrm{Disc}}$ and $L_{\mathrm{Power}}$ are the unabsorbed disc luminosity in $\mathrm{0.001-100.0 \ keV}$ and unabsorbed powerlaw luminosity in $\mathrm{1.0-100.0 \ keV}$ energy range, respectively. We used the convolution model \texttt{CFLUX} in \texttt{XSPEC} for flux calculations of each component.

\begin{table}
\centering
	\caption{Values adopted for model parameters of \texttt{DISKBB}, \texttt{KERRBB} and \texttt{KYNBB}} 
	\label{table:model_pars}
	\begin{tabular}{p{0.14\textwidth}p{0.14\textwidth}p{0.14\textwidth}}
		\hline \hline
		 Model Parameter & Value & Free/Frozen \\
		 \hline
		&\texttt{DISKBB}\\
		\hline
		$\rm{T}_{\mathrm{in}}$ & -- & Free \\
		Normalisation & -- & Free \\
		
		\hline
		&\texttt{KERRBB}\\
		\hline

        $\eta$ & 0 & Frozen \\
		Black Hole Spin & -- & Free (see Section~\ref{spectral_analysis}) \\
		Inclination Angle & (see Table~\ref{table:src}) & Frozen \\
		Black Hole Mass & (see Table~\ref{table:src}) & Frozen \\
		Mass Accretion Rate & -- & Free \\
		Distance & (see Table~\ref{table:src}) & Frozen \\
		$f_{\mathrm{col}}$ & 1.7 & Frozen (see Section~\ref{f_col}) \\
        Self-irradiation Flag & 0 & Frozen \\
        Limb-darkening Flag & 0 & Frozen \\
        Normalisation & 1 & Frozen\\

		\hline
		&\texttt{KYNBB}\\
		\hline
		Black Hole Spin & -- & Free (see Section \ref{spectral_analysis}) \\
		Inclination Angle & (see Table \ref{table:src}) & Frozen \\
		Inner Radius & -- & Free \\
		Switch for Inner Radius & 0 & Frozen \\
		Outer Radius & 1000 & Frozen \\
		$\phi$ & 0 & Frozen \\
		$d\phi$ & 360 & Frozen \\
		Black Hole Mass & (see Table \ref{table:src}) & Frozen \\
		Mass Accretion Rate & -- & Free \\
		$f_{\mathrm{col}}$ & 1.7 & Frozen \\
		$\alpha$ & 0 & Frozen \\
		$\beta$ & 0 & Frozen \\
		$r_{\mathrm{cloud}}$ & 0 & Frozen \\
		Redshift & 0 & Frozen \\
		ntable & 80 & Frozen\\
		nrad & 150 & Frozen\\
		Division & 1 & Frozen \\
		$n\phi$ & 180 & Frozen \\
		Smooth & 0 & Frozen \\
		Stokes & 0 & Frozen \\
		$\chi_{0}$ & 0 & Frozen \\
		$\tau$ & 11 & Frozen \\
		nthreads & & Frozen\\
		Normalisation & (see Table \ref{table:src} and Equation \ref{normalization-kynbb}) & Frozen\\
		
		\hline
		
		\hline \hline
		
	\end{tabular}
\end{table}

\section{RESULTS}\label{results}
\subsection{GRO J1655-40}

We first analysed the entire sample of observations of GRO J1655-40 listed in Table~\ref{table:obs} using \texttt{DISKBB}. The majority of the excluded observations with reduced $\chi^{2}>2.0$ were distributed throughout the entire outburst with a subset concentrated in the region of the HID corresponding to the transition from SIMS to HSS when tracked on a light curve and an HID. Spectral fitting of these specific observations with \texttt{DISKBB} showed residuals at energies between 5-7 keV. 

We proceeded further to obtain the flux values of the unabsorbed model components separately. Due to unconstrained parameters of the disc model, the fitting procedure with \texttt{PyXspec} during flux calculations of the disc component was interrupted for states with significantly low count rates where the source was either at the very beginning or the end of the outburst. A total of 71 observations are excluded from further analysis due to this complication. We also obtained significantly high reduced $\chi^{2}$ values for the observations covering the outbursts from 1996 and 1997 due to a significant increase in the sensitivity of the PCA instrument after calibration with \texttt{pcacorr} instead of previously adopted methods. Simple spectral modelling with \texttt{DISKBB} with just a powerlaw component produced significant residuals for some of these observations at energies between 5-7 keV which couldn't be modelled with a simple Gaussian line (\texttt{GAUSS}) with line energies between 6-7 keV. While an additional \texttt{GAUSS} component improved the fits significantly, these observations did not fall into our selection criteria due to reduced $\chi^{2}$ values well above 10, suggesting a more detailed analysis of these specific observations. More sophisticated modelling of the reflection component introduces a new level of model dependency when used with \texttt{KERRBB} and \texttt{KYNBB} since both models still extend to higher energies, eventually coinciding with the 5-7 keV energy range. With RXTE's limited spectral resolution, proper modelling of such line features introduces a risk of bias we choose to avoid for the purpose of our analysis. A detailed analysis of the line properties will be presented in further studies. Hence, we exclude these 81 observations from these two outbursts for further investigation and only include the 2005 outburst which already encompasses more than 86\% of the entire sample for GRO J1655-40. The number of remaining observations after the selection criteria applied for GRO J1655-40 stood at 202 for \texttt{DISKBB}.

Using the reduced set of observations, we replaced \texttt{DISKBB} with \texttt{KERRBB}. We also tested the poor fits obtained in the previous step of the analysis with \texttt{KERRBB} but did not observe any significant improvement in reduced $\chi^{2}$ values to match the selection criteria. Throughout our analysis with \texttt{KERRBB}, we left the mass accretion rate $\dot{M}$ as a free parameter and use a fixed colour correction factor of $f_{\mathrm{col}}=1.7$ throughout the outbursts \citep{Shimura95}. We also fixed $\eta = 0$ to adopt a stress-free boundary condition at $R_{\rm ISCO}$. Following the discussion in \citet{Li05}, we fix the normalisation at 1.0 since the mass, inclination angle of the disc and distance to the source are fixed while one still has the freedom of choosing a different value due to the uncertainties in measurements of these parameters.

We first attempted to fit the same set of spectra with a fixed spin at $a_{*}=0.7$ following the results from the X-ray spectral continuum fitting method by \citet{Shafee06}. Adopting a spin value fixed at a value in \texttt{KERRBB} produced poorer fits with reduced $\chi^{2}$ much larger than 2.0 for $\sim$ 89\% of observations due to the residuals in the lower energies where the model was not able to reproduce the spectral shape for the same physical parameters and a specific spin value. As a result, we carried out the analysis for GRO J1655-40 with a spin parameter left free to vary between 0.0-0.9982. Figure~\ref{fig:r_in-time-gro} shows the evolution of the measured spin parameter throughout the outburst. Limiting the spin parameter to a smaller range between 0.65-0.75 following the measurements by \citet{Shafee06} did not improve the number of fits with reduced $\chi^{2}<2.0$. 

In contrast to \texttt{DISKBB}, disc temperature is not a model parameter in \texttt{KERRBB} and \texttt{KYNBB}. To obtain the value for the disc temperature in a similar sense to \texttt{DISKBB}, we evaluated the position of the peak of the disc component in the model. Due to the zero-torque boundary condition at the innermost disc radius adopted by \texttt{KERRBB}, this peak corresponds to a disc radius not exactly at ${R}_{\mathrm{ISCO}}$ but slightly further out. This is still a valid approximation of the accretion disc temperature as measured by \texttt{KERRBB}.

\begin{figure}
\includegraphics[width=0.47\textwidth]{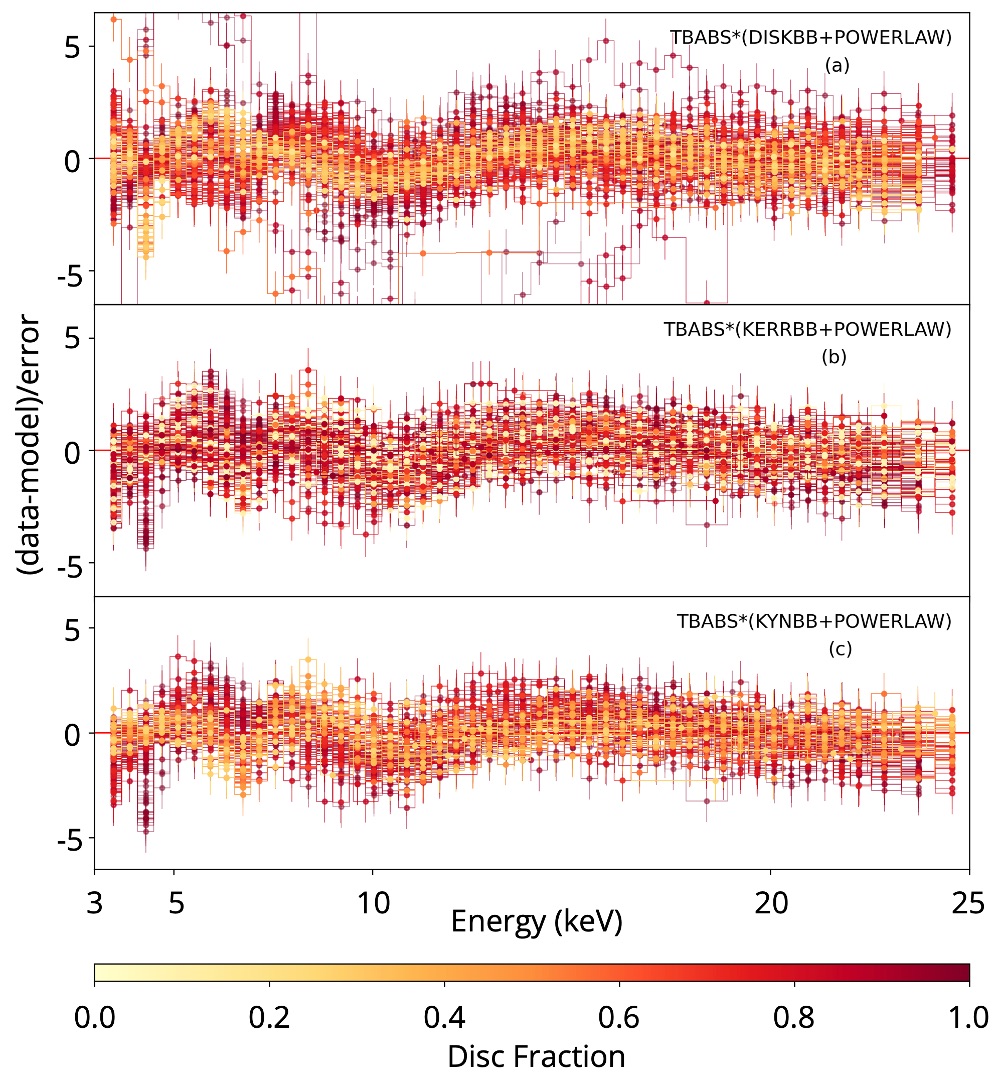}
\caption{(a) Residuals in (data-model)/error for fits of GRO J1655-40 with reduced $\chi^{2} < 2.0$ obtained from spectral fitting with \texttt{TBABS$\times$(DISKBB+POWERLAW)}
(b) Residuals in (data-model)/error for fits of GRO J1655-40 with reduced $\chi^{2} < 2.0$ obtained from spectral fitting with \texttt{TBABS$\times$(KERRBB+POWERLAW)}
(c) Residuals in (data-model)/error for fits of GRO J1655-40 with reduced $\chi^{2} < 2.0$ obtained from spectral fitting with \texttt{TBABS$\times$(KYNBB+POWERLAW)}}
    \label{fig:spec-gro-delchi}
    \textit{} 
\end{figure}

Due to difficulties in fitting with a fixed spin for GRO J1655-40, we replaced the relativistic disc model with \texttt{KYNBB} to have more freedom with the definition of the innermost edge of the accretion disc $R_{\rm in}$. Additional to the mass accretion rate, we let $R_{\rm in}$ vary throughout our analysis with \texttt{KYNBB}. This allowed us to recover 93\% of the observations \texttt{KERRBB} was able to fit only with a free spin parameter. We fixed the normalisation parameter to their respective values following  equation~(\ref{normalization-kynbb}) calculated using distances adopted in Table.
We present the residuals in (data-model)/error obtained from our spectral fits of GRO J1655-40 for each model in Figure~\ref{fig:spec-gro-delchi}.

Apart from the initial sample obtained by fitting with \texttt{DISKBB}, we also analysed the 1996 and 1997 outbursts of GRO J1655-40 with \texttt{KERRBB} but we observed no significant improvement in reduced $\chi^{2}$ below 2. Due to the long computation times required for \texttt{KYNBB}, we did not proceed with the same analysis with \texttt{KYNBB} for these outbursts. We present the spectral fits of these outbursts with \texttt{DISKBB} and \texttt{KERRBB} in Figure~\ref{fig:spec-gro-96-97} in Appendix A.

\subsection{LMC X-3}
We adopted the same approach to LMC X-3 and first analysed the entire sample of observations with \texttt{DISKBB}. We applied the same selection criteria with a cut at reduced $\chi^{2} = 2.0$ excluding the observations that lie outside this range. We also excluded the observations where the disc component was not constrained and caused the fitting procedure with \texttt{PyXspec} to be interrupted during the flux calculations with \texttt{CFLUX}. The number of such observations was significantly lower due to the almost-persistent nature of LMC X-3 in contrast to GRO J1655-40.

We continued with our analysis using \texttt{KERRBB} for LMC X-3 to investigate the behaviour of the spin parameter across different luminosities. We fixed the physical parameters to their respective values listed in Table~\ref{table:src}. Unlike in the case of GRO J1655-40, we were able to fit $\sim$ 95\% of these observations with a spin fixed at $a_{*}=0.25$ following \citet{Steiner14}. The outliers in this sample did not correspond to a specific state of the system, showing no indication of the effects of a significant change in the inner disc radius affecting the fitting process as observed in GRO J1655-40.

\begin{figure}
\includegraphics[width=0.47\textwidth]{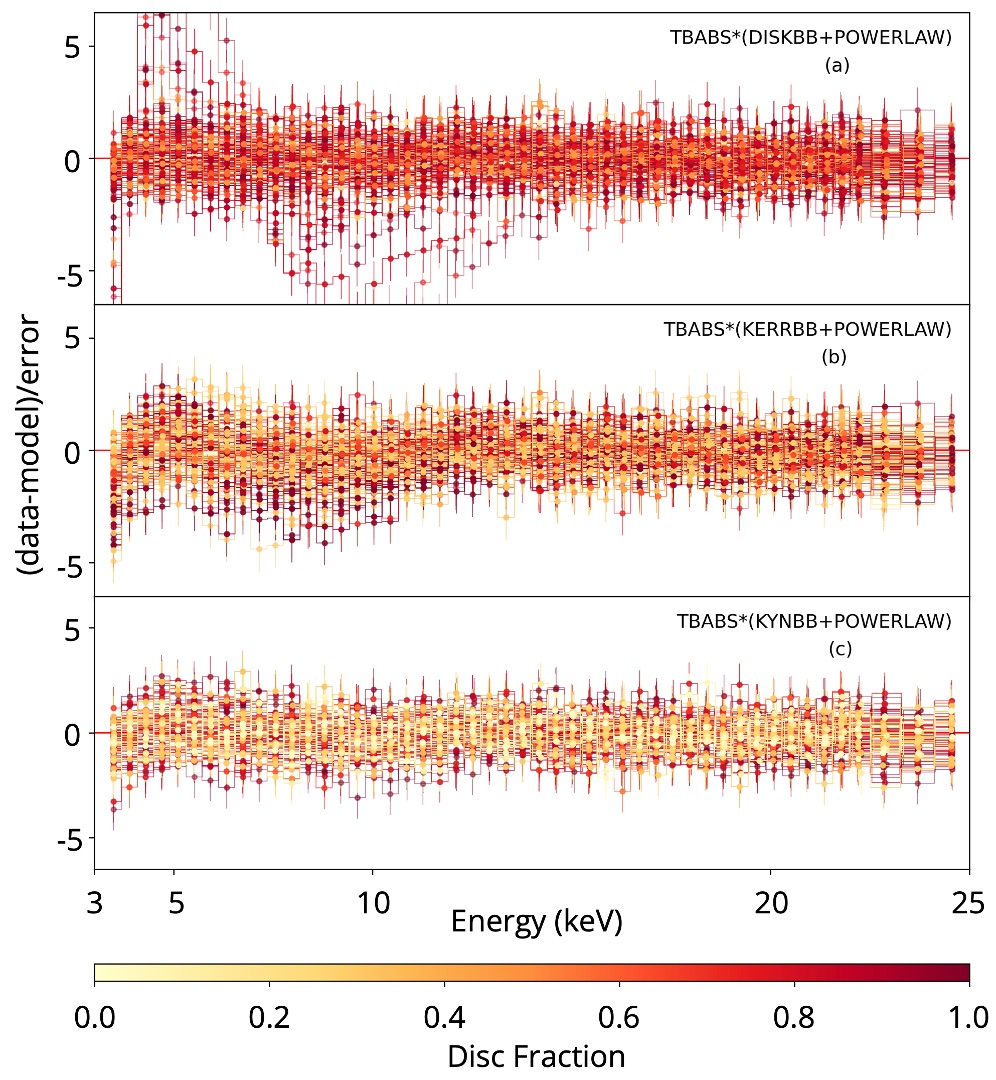}
\caption{(a) (data-model)/error residuals for fits of LMC X-3  with reduced $\chi^{2} < 2.0$ obtained from spectral fitting with \texttt{TBABS$\times$(DISKBB+POWERLAW)} for observations with observation ID's between 80103-01-24-00 and 91105-03-03-11.
(b) (data-model)/error residuals for fits of LMC X-3 with reduced $\chi^{2} < 2.0$ obtained from spectral fitting with \texttt{TBABS$\times$(KERRBB+POWERLAW)} for observations with observation ID's between 80103-01-24-00 and 91105-03-03-11.
(c) (data-model)/error residuals for fits of LMC X-3 with reduced $\chi^{2} < 2.0$ obtained from spectral fitting with \texttt{TBABS$\times$(KYNBB+POWERLAW)} for observations with observation ID's between 80103-01-24-00 and 91105-03-03-11.} 
    \label{fig:spec-lmc-delchi}
    \textit{} 
\end{figure}

Following our analysis of GRO J1655-40, we used the same setup for \texttt{KYNBB} and fit the spectra with a fixed spin at $a_{*}=0.25$ and values for the black hole mass, inclination angle, the distance to the source (i.e. the normalisation parameter in \texttt{KYNBB}) to their respective values and did not change any of the other model parameter setups we used for GRO J1655-40. We present the residuals in (data-model)/error obtained from our spectral fits of LMC X-3 for each model in Figure~\ref{fig:spec-lmc-delchi}.

\begin{figure}
\centering
\includegraphics[width=0.49\textwidth]{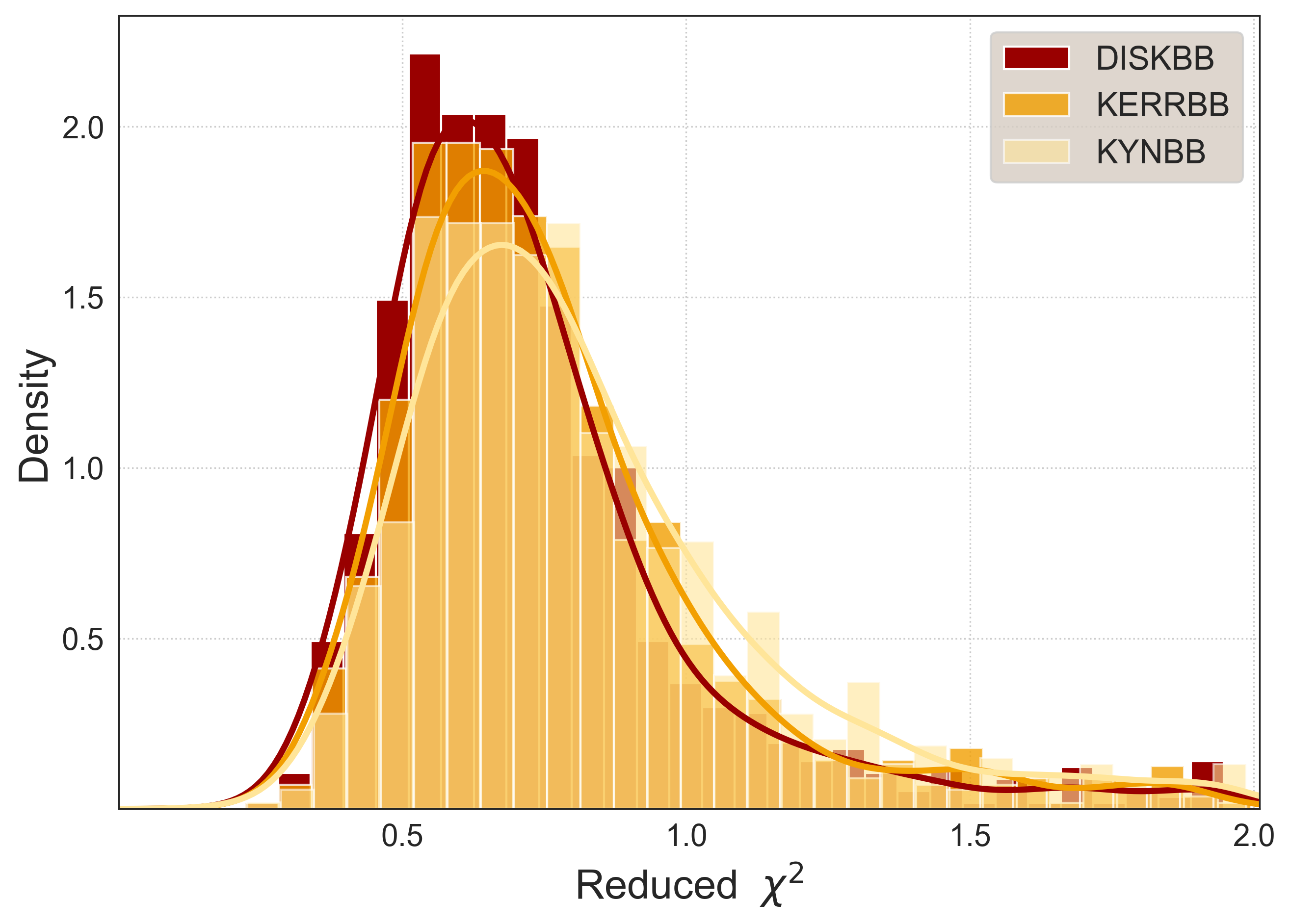}
\includegraphics[width=0.49\textwidth]{plots/chi-all-freq.png}
\caption{The distribution of reduced $\chi^{2}$ values obtained from all of the fits with $\chi^{2}/\rm{d.o.f.}<2.0$ analysed with \texttt{DISKBB} (red), \texttt{KERRBB} (orange) and \texttt{KYNBB} (yellow) for GRO J1655-40 and LMC X-3 combined.}
    \label{fig:red-chi-gro}
\end{figure}

In Figure~\ref{fig:red-chi-gro}, we present the distribution of reduced $\chi^{2}$ values from all of the fits with $\chi^{2}/\rm{d.o.f.}<2.0$ across each model for GRO J1655-40 and LMC X-3 combined. With each fit, the number of observations producing good fits was reduced with a cut at $\chi^{2}/\mathrm{d.o.f.}<2$. For a big majority of these observations, adding a simple Gaussian line did not recover the fit. To preserve the simplicity of the model configuration described in Section~\ref{spectral_analysis}, we did not perform a more sophisticated analysis for these specific observations and instead chose to remove them from our analysis. 

\subsection{Disc Luminosity ${L_{\rm {Disc}}}$ and Disc Temperature ${T_{\rm {Disc}}}$}\label{t_lumin}

\begin{figure}
\centering
\includegraphics[width=0.49\textwidth]{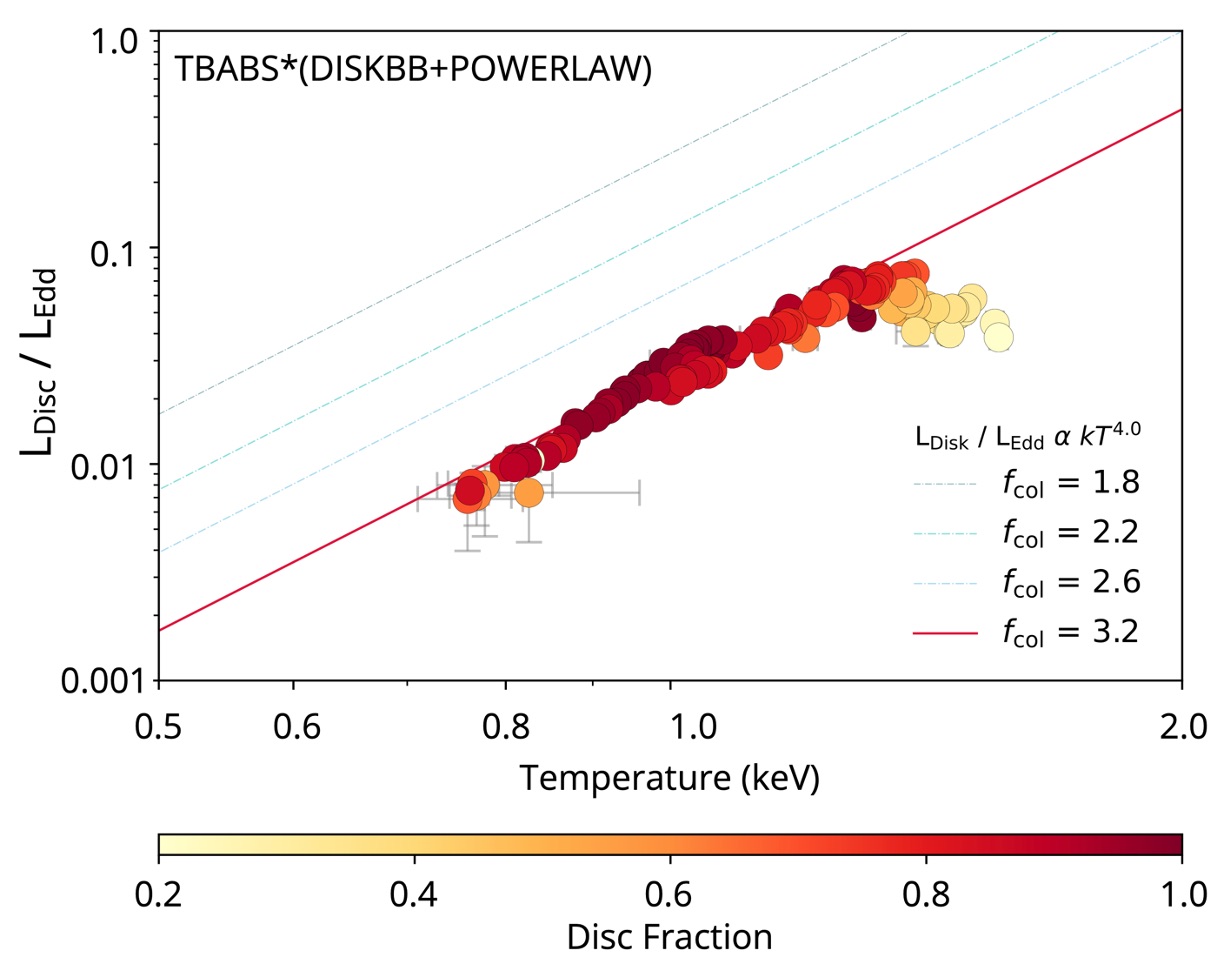}
\includegraphics[width=0.49\textwidth]{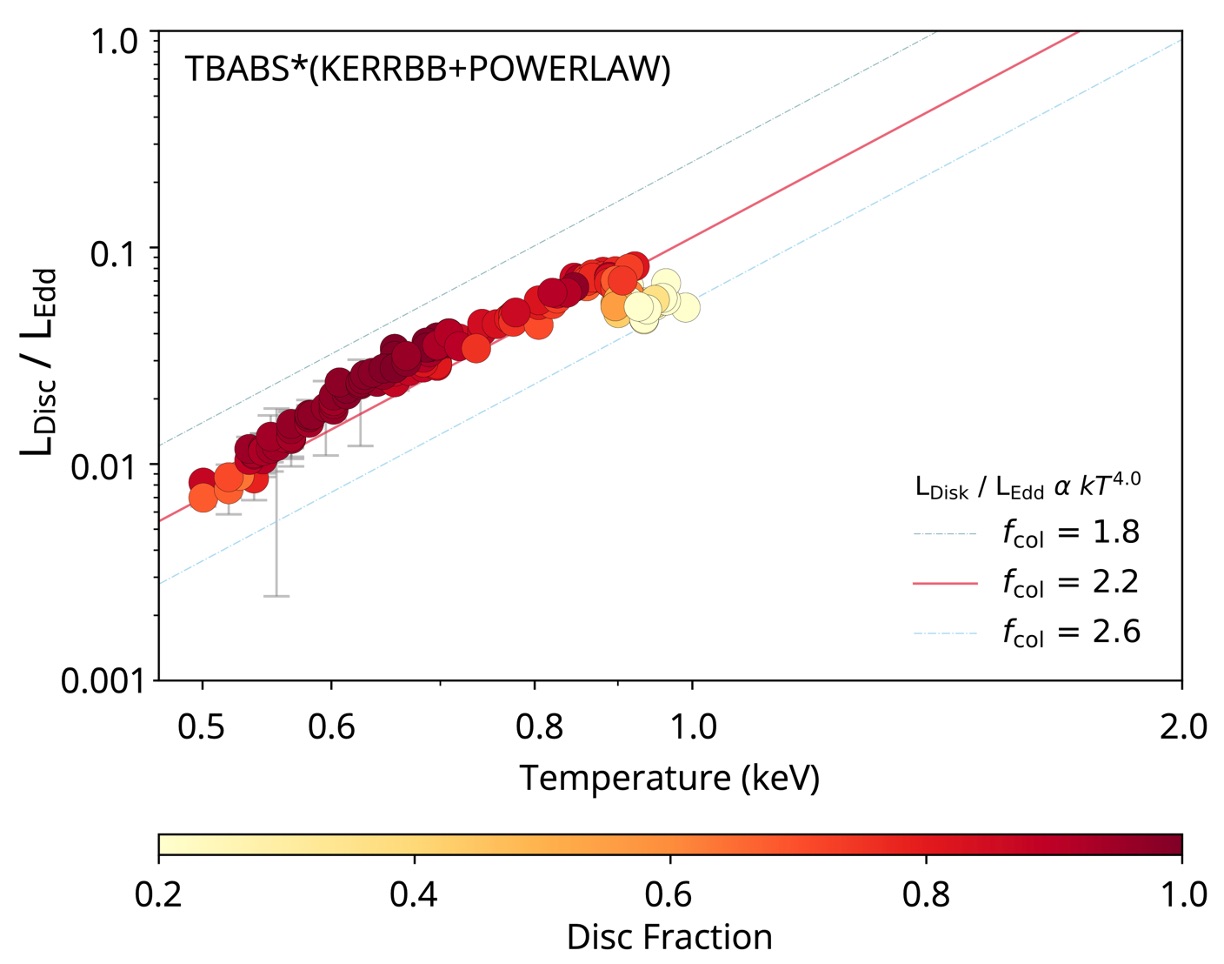}
\includegraphics[width=0.49\textwidth]{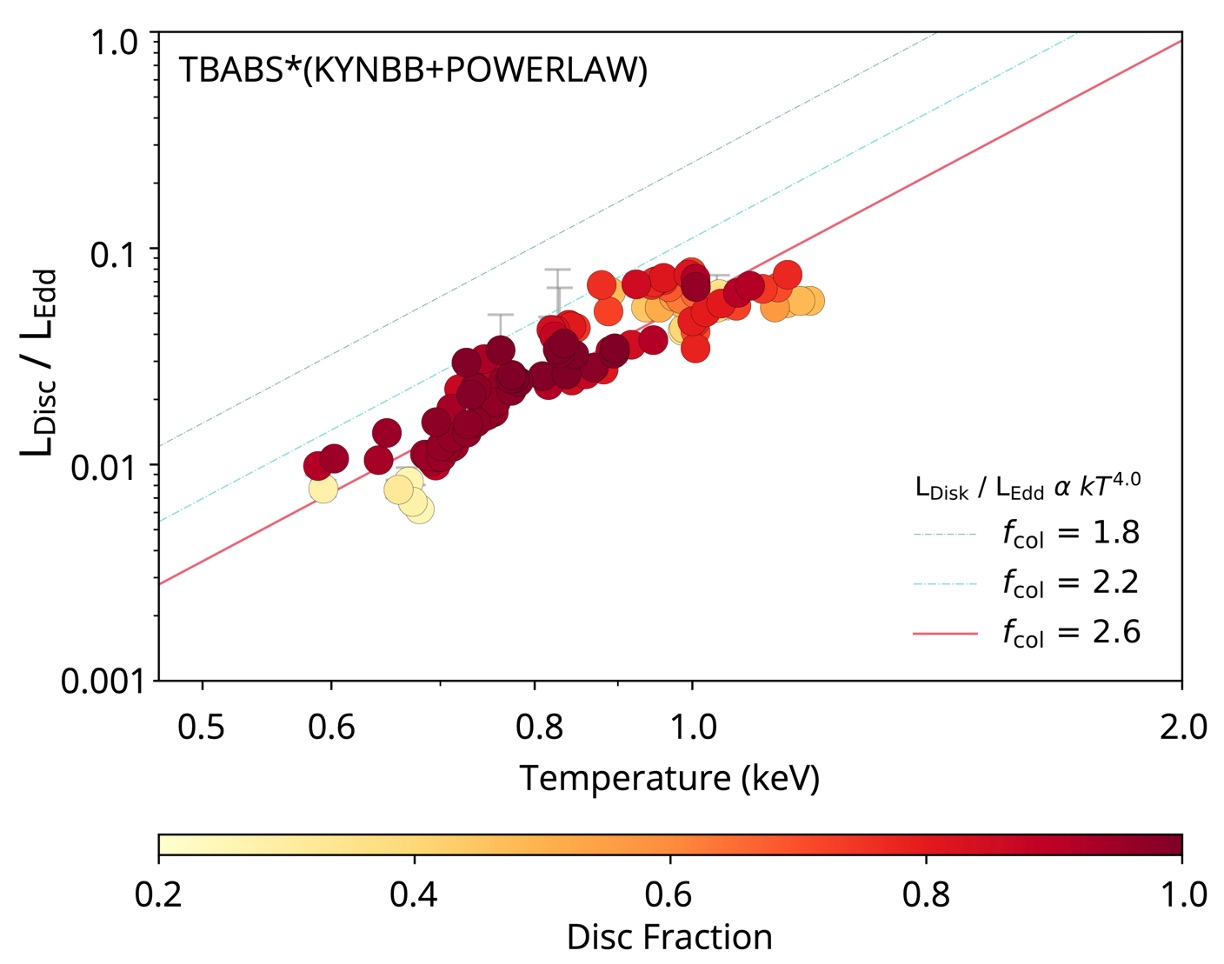}
\caption{\textit{Top}: Change in disc luminosity with respect to  disc temperature for GRO J1655-40 as measured by \texttt{DISKBB}, \textit{Middle}: Change in disc luminosity with respect to  disc temperature for GRO J1655-40 as measured by \texttt{KERRBB}, \textit{Bottom}: Change in disc luminosity with respect to  disc temperature for GRO J1655-40 as measured by \texttt{KYNBB}}.
    \label{fig:temp-edd-gro}
    \textit{} 
\end{figure}

We used the physical description from \citet{Gierlinski99} to calculate the relation between the disc luminosity and the temperature,
\begin{equation}\label{eddington_ratio}
\frac{L_{\text {Disc }}}{L_{\text {Edd}}} \approx 0.583\left(\frac{1.8}{f_{\text {col }}}\right)^{4}\left(\frac{M}{10 \mathrm{M}_{\odot}}\right)\left(\frac{k T_{\text {max }}}{1 \mathrm{keV}}\right)^{4}
\end{equation}
Following the general description for \texttt{DISKBB}, to account for strong gravitational potentials, we adopt the following relativistic correction by \citet{Zhang97} between $T_{\mathrm{max}}$ and $T_{\mathrm{obs}}$

\begin{equation}\label{t_max}
T_{\rm max }=T_{\rm obs} / f_{G R}(\theta, a_{*}) \xi
\end{equation}
where $\xi=1.04$ is for the stress-free boundary layer, $f_{G R}(\theta, a_{*})$ is the fractional change of the colour temperature (assumed 1.064 following Table 1 in \citet{Zhang97}, derived by \citet{Cunningham75}), $\theta$ is the inclination angle and $a_{*}$ the dimensionless spin parameter. Dotted lines in Figure~\ref{fig:temp-edd-gro} show the calculated relationships for GRO J1655-40 for different $f_{\mathrm{col}}$ values in comparison to the observed trend which is indicated with a red solid line.

We obtained similar results as previously presented  by \citet{Gier04}, \citet{Done07} and \citet{Dunn08,Dunn10, Dunn11} from the fitting with \texttt{DISKBB} for GRO J1655-40 and LMC X-3. 2005 outburst of GRO J1655-40, analysed with \texttt{DISKBB}, revealed the same deviations from expected the $L_{\rm Disc}$ - $T_{\rm Disc}$ trend or "spurs" as previously labeled by \citet{Dunn11}. We present the changes in disc luminosities with respect to disc temperatures ($L_{\rm Disc}$ - $T_{\rm Disc}$) obtained from \texttt{DISKBB}, \texttt{KERRBB} and \texttt{KYNBB} for $f_{\mathrm{col}} = 1.7$ in Figure~\ref{fig:temp-edd-gro}. These structures correspond to observations where the luminosity remains comparatively the same while the temperature rises significantly, making these specific states shift towards the right of the general trend. Since these deviations from the trend are observed in the states corresponding to the smallest inner disc radii and highest disc temperature, we explored the possible effect of changes in $f_{\mathrm{col}}$ due to changes in the accretion states by letting $f_{\mathrm{col}}$ as a free parameter in \texttt{KERRBB} throughout the outburst. Our results did not indicate any changes to the observed trend in the upper and middle panels of Figure~\ref{fig:temp-edd-gro}. As also presented by \citet{Dunn11}, we did not observe these deviations from the general trend for the entire sample of LMC X-3 observations. We, therefore, do not include the corresponding plots for LMC X-3 in further discussion.

For the entire set of observations of both GRO J1655-40 and LMC X-3, we measured $\sim$52\% (on average) higher disc temperatures with \texttt{DISKBB} than \texttt{KERRBB} while $L_{\rm Disc}$ is measured to be $\sim$8\% lower on average. This temperature difference, however, did not provide an improvement to the previously reported "spurs" since this temperature change was a result of $f_{\mathrm{col}}$ correction to the spectrum and was applied throughout the entire sample during the analysis, not for these states only. When compared to \texttt{KYNBB} with the inner disc radius as a free parameter, \texttt{DISKBB} gives disc temperature values 40\% higher on average while the luminosity is measured to be 2\% lower. When the two relativistic accretion disc models are compared, \texttt{KYNBB} gives disc temperatures 18\% higher than \texttt{KERRBB} on average while the disc luminosities are very similar with \texttt{KYNBB} measuring only 0.3\% higher on average than \texttt{KERRBB}. 

Figure~\ref{fig:temp-edd-gro} shows that an initial comparison of the non-relativistic disc model \texttt{DISKBB} to its relativistic counterpart \texttt{KERRBB} was inconclusive in the context of treating the trend in $L_{\rm Disc}$ - $T_{\rm Disc}$ previously reported by \citet{Dunn10}. Adopting a relativistic model where the inner disc radius was left free to vary with \texttt{KYNBB} provided a significant recovery to the $L_{\rm Disc}$ - $T_{\rm Disc}$ trend observed by \texttt{DISKBB} and \texttt{KERRBB} for GRO J1655-40 while the disc temperature for the rest of the sample remained dominantly unchanged with only 3\% of increase on average. The observations corresponding to these observed deviations were moved to lower temperatures matching other observations with similar $L_{\rm Disc} / L_{\rm Edd}$ values. We also observed more scattering in the trend possibly due to $R_{\rm in}$ being an additional free parameter, causing larger deviations in the scatter compared to \texttt{DISKBB} and \texttt{KERRBB}.

\subsection{Inner Disc Radius $R_{\mathrm{in}}$ and Disc Temperature $T_{\mathrm{in}}$}\label{radius}
\begin{figure}
\centering
\includegraphics[width=0.49\textwidth]{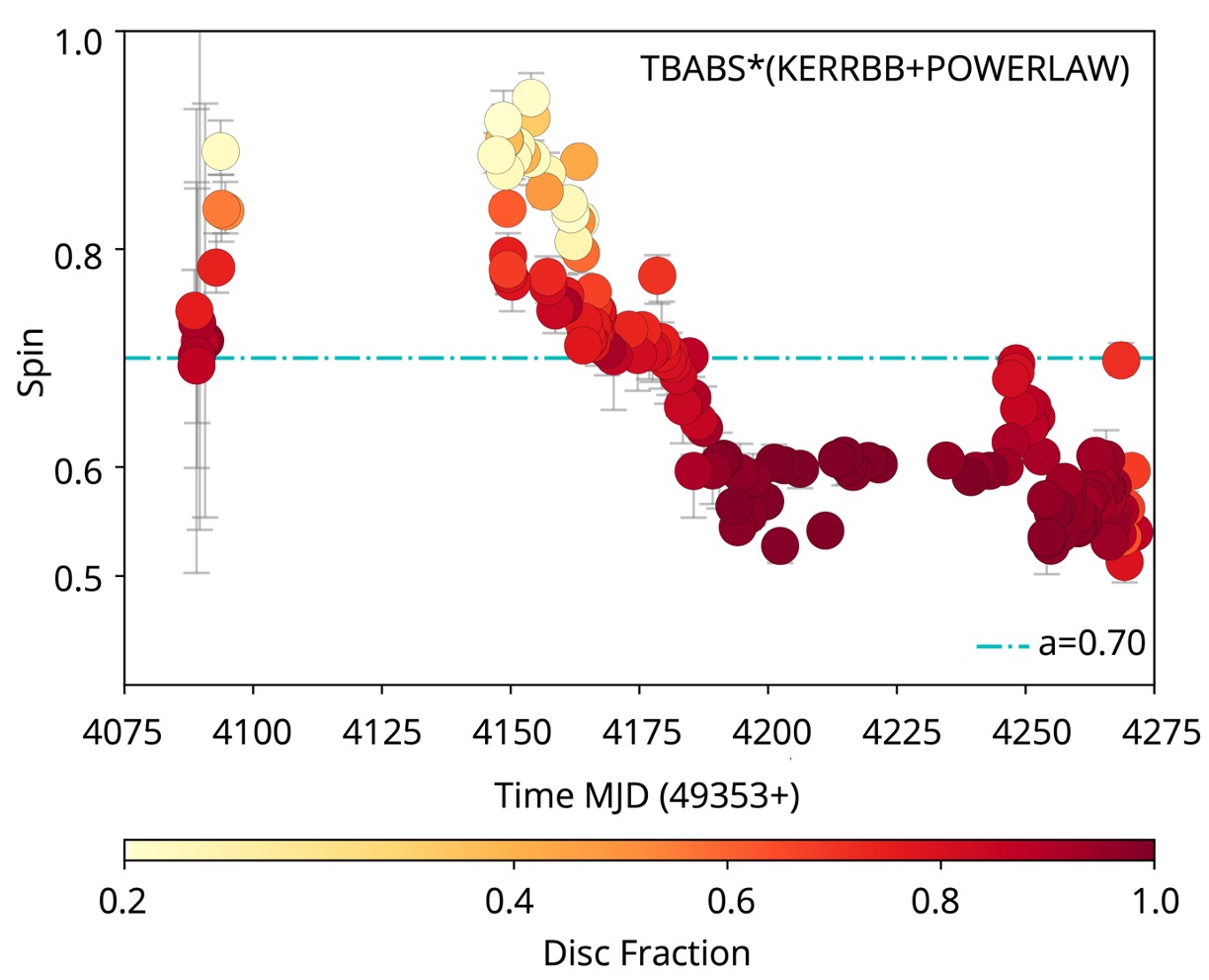}
\includegraphics[width=0.49\textwidth]{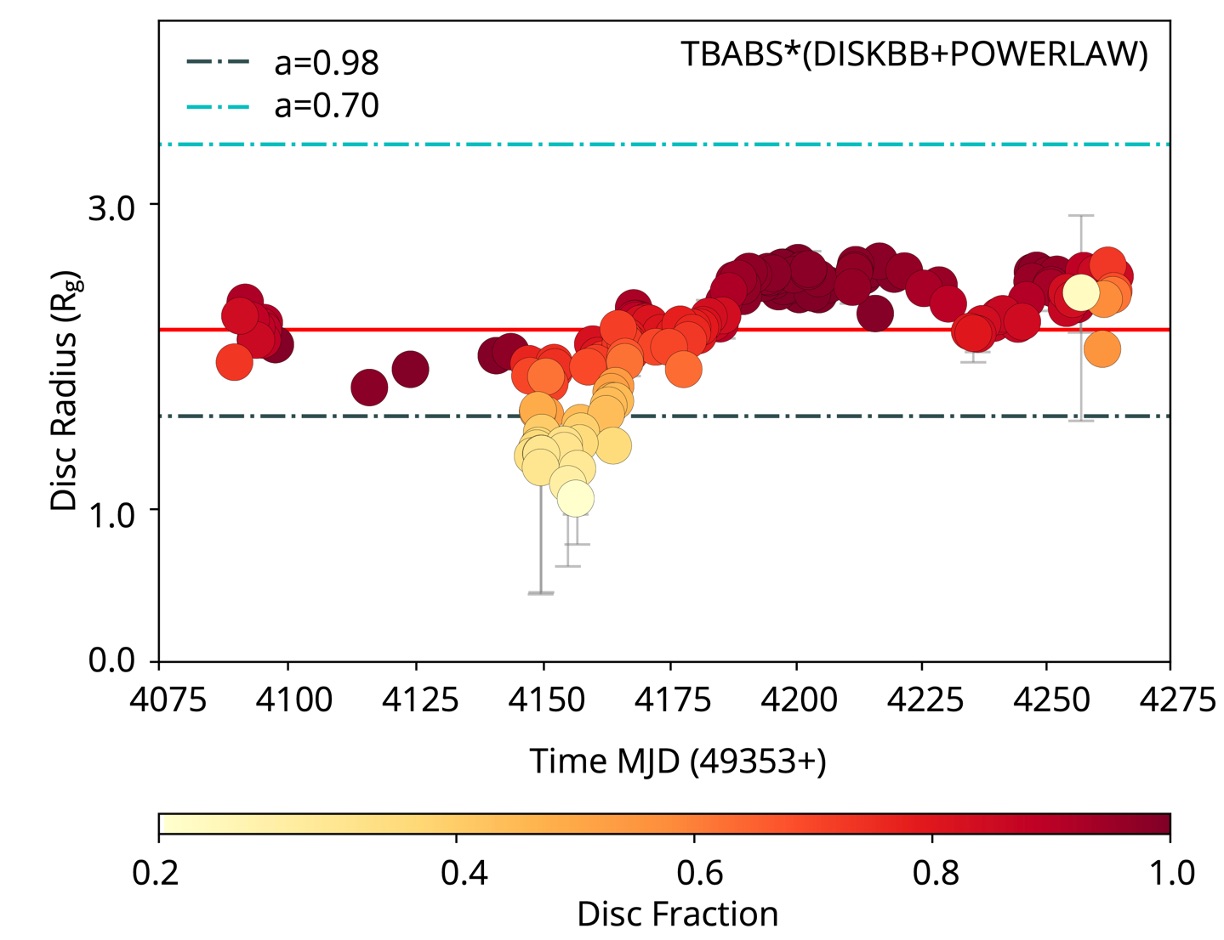} 
\includegraphics[width=0.49\textwidth]{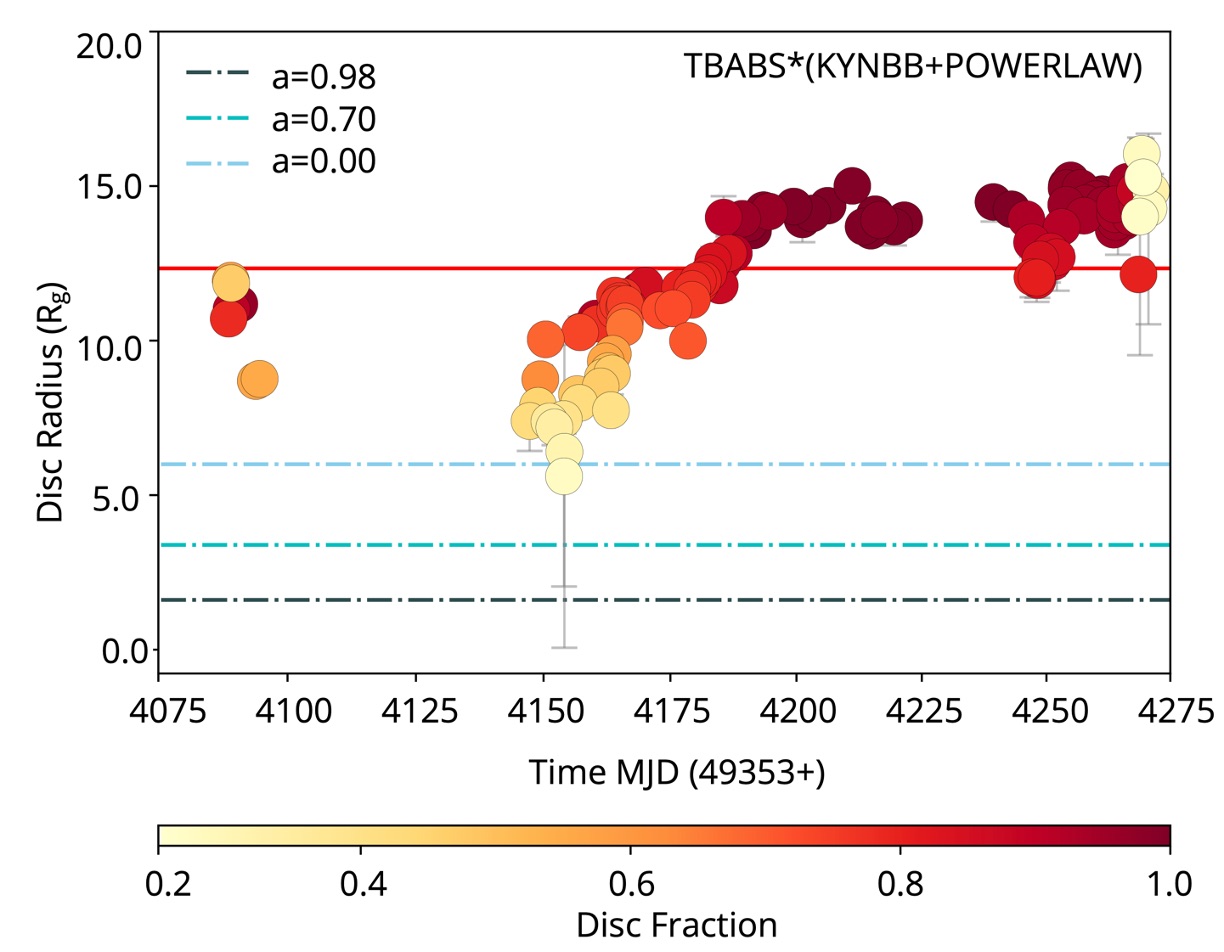}
\caption{\textit{Top}: Change in the spin measurements with respect to time for GRO J1655-40 as measured by \texttt{KERRBB}, \textit{Middle}: change in the disc radius measurements with respect to time as measured by \texttt{DISKBB} in units of $\rm R_{g}$, \textit{Bottom}: Change in the disc radius measurements with respect to time for GRO J1655-40 as measured by \texttt{KYNBB} in units of $\rm R_{g}$. Red vertical solid lines in the middle and bottom plots indicate the average of the measured inner disc radius values.}
    \label{fig:r_in-time-gro}
    \textit{} 
\end{figure}

Our first attempt at spectral analysis for GRO J1655-40 using \texttt{KERRBB} with a fixed spin produced significantly high $\chi^{2}/\rm{d.o.f.}$ values for $\sim$ 89\% of our sample. We didn't find a certain spin value the fit favoured significantly in terms of the goodness of the fit with the model. We explored the parameter space of \texttt{KERRBB} to better understand the nature of the source that would mimic a changing spin scenario by first attempting to fix the spin parameter at 0.7 following \citet{Shafee06, Stuchl16} and leave the normalisation free throughout the sample. This allowed us to have a global spin value for the sample but revealed a big variation in the normalisation values. The normalisation parameter of \texttt{KERRBB} is sensitive to uncertainties in the black hole mass, distance to the source and the inclination angle of the system. However, such a great variation in the normalisation cannot be explained by uncertainties in these parameters of the model and shouldn't be different on this scale for each observation. Considering the core assumption of \texttt{KERRBB} that the position of $R_{\mathrm{in}}$ always terminates at $R_{\mathrm{ISCO}}$ and dependence of $R_{\mathrm{ISCO}}$ on the black hole mass in  equation~(\ref{r_isco}), we interpret this as a strong indication for changing inner disc radius throughout the outburst. As a result, our attempt to fit the entire outburst with a fixed global spin parameter with \texttt{KERRBB} produced significantly poorer fits compared to different combinations of free parameters we applied before. We then analysed the same set of spectra with \texttt{KYNBB} replacing \texttt{KERRBB} with $R_{\mathrm{in}}$ as a free parameter. We found the same trend in the evolution of the inner disc radius both with \texttt{DISKBB} and \texttt{KYNBB} and the spin parameter measured by \texttt{KERRBB} showed to be in the expected reverse trend throughout the outburst. We present this change in the measured dimensionless spin parameter with respect to time in the top panel of Figure~\ref{fig:r_in-time-gro} in comparison with the change in disc radius measured by \texttt{DISKBB} and \texttt{KYNBB} (middle and bottom panels, respectively).

\begin{figure}
\centering
\includegraphics[width=0.49\textwidth]{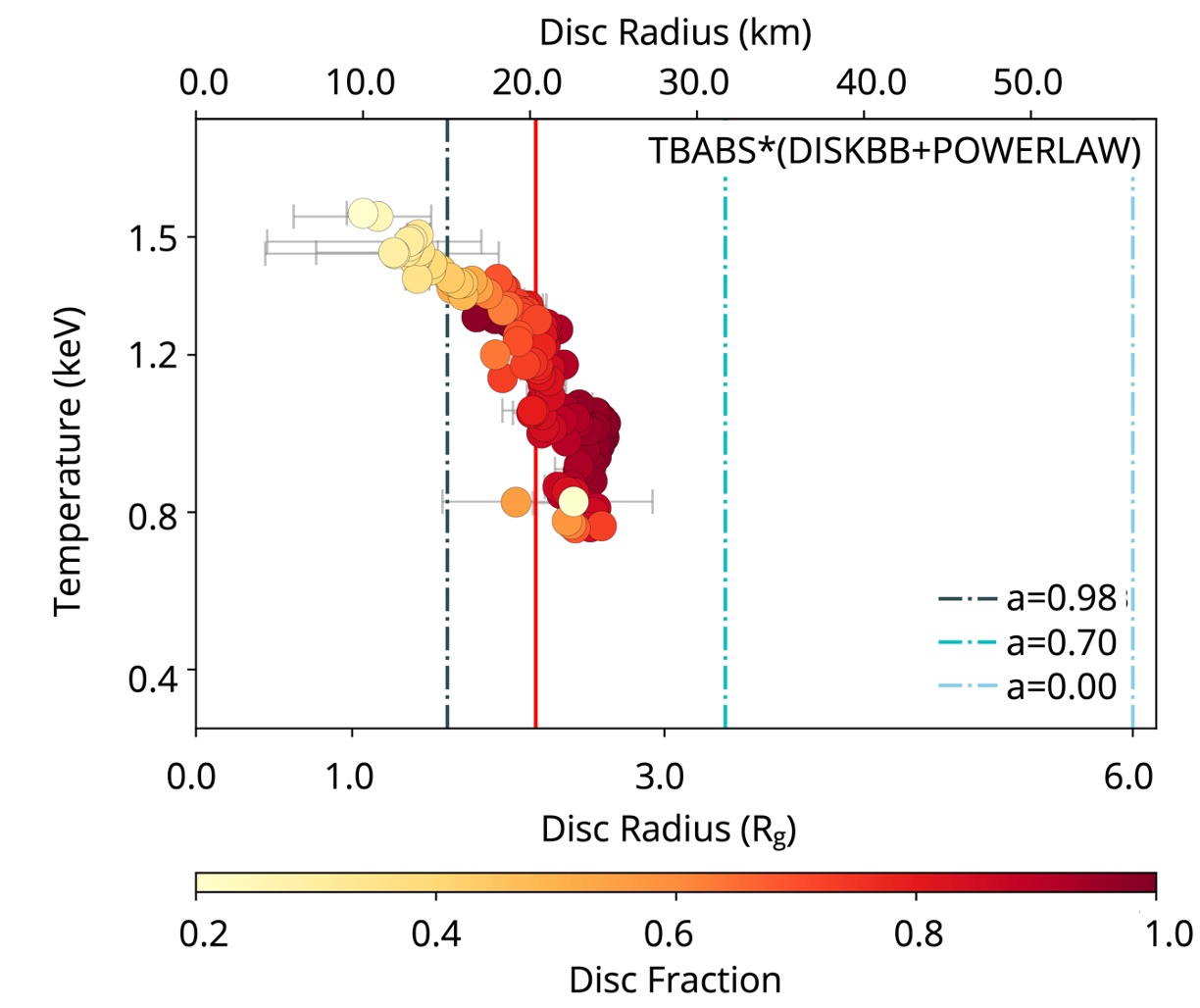}
\includegraphics[width=0.49\textwidth]{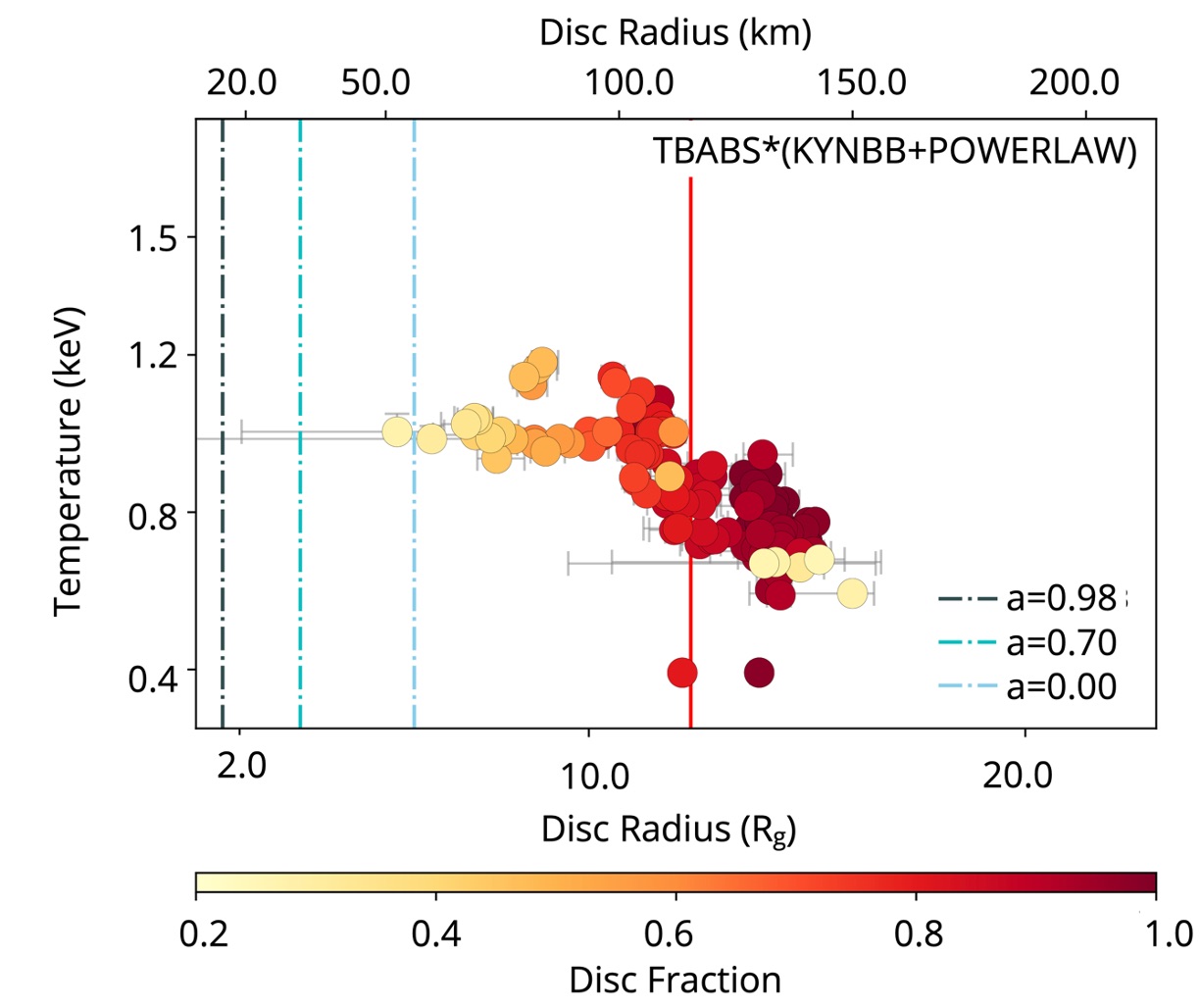}
\caption{\textit{Top}: Relationship between the inner disc radii and the disc temperature for GRO J1655-40 as measured by \texttt{DISKBB} in units of $\rm R_{g}$ \textit{Bottom}: Relationship between the inner disc radii and the disc temperature for GRO J1655-40 as measured by \texttt{KYNBB} in units of $\rm R_{g}$. Dashed lines correspond to $\rm R_{ISCO}$ values for the spin parameter at 0.0, 0.7 and 0.98, respectively. Red vertical solid lines indicate the average of the measured inner disc radius values.}
    \label{fig:r_in-temp_gro}
    \textit{} 
\end{figure}

We also investigated the relationship between the inner disc radius and the disc temperature using \texttt{DISKBB} and \texttt{KYNBB}, comparing these results for both non-relativistic and relativistic accretion disc models. As expected, our results showed that $R_{\mathrm{in}}$ measurements calculated from the normalisation of \texttt{DISKBB} correspond to physically unrealistic values that are way beyond the $R_{\mathrm{ISCO}}$ even for a Kerr black hole with $a_{*} = 0.98$ (Figure~\ref{fig:r_in-temp_gro}) for the big majority of the observations. It is important to note that this method of inner disc radius calculation is still affected by uncertainties in measurements of the inclination angle of the disc and distance to the source. However, subtle changes in these parameters aren't expected to produce such small values for $R_{\mathrm{in}}$ in comparison with $R_{\mathrm{ISCO}}$ values calculated for different spin parameters, occasionally reaching a scale of having an accretion disc with its inner edge extending down to the event horizon for a Schwarzschild black hole.  

Our analysis with \texttt{KYNBB}, however, revealed much larger radii corresponding well above $R_{\mathrm{ISCO}}$ for $a_{*} = 0.7$. We observed a very similar pattern for the changes in the inner disc radius to the results from \texttt{DISKBB} but with values that are at least 5 times larger on average. Except for two observations, all measurements corresponded to regions with $R_{\mathrm{in}} > R_{\mathrm{ISCO}}$ for all spin values. In addition to significantly larger values, these results also cover a much wider parameter range (50-150 km) compared to \texttt{DISKBB} (10-25 km). 

\begin{figure}
\centering
\includegraphics[width=0.49\textwidth]{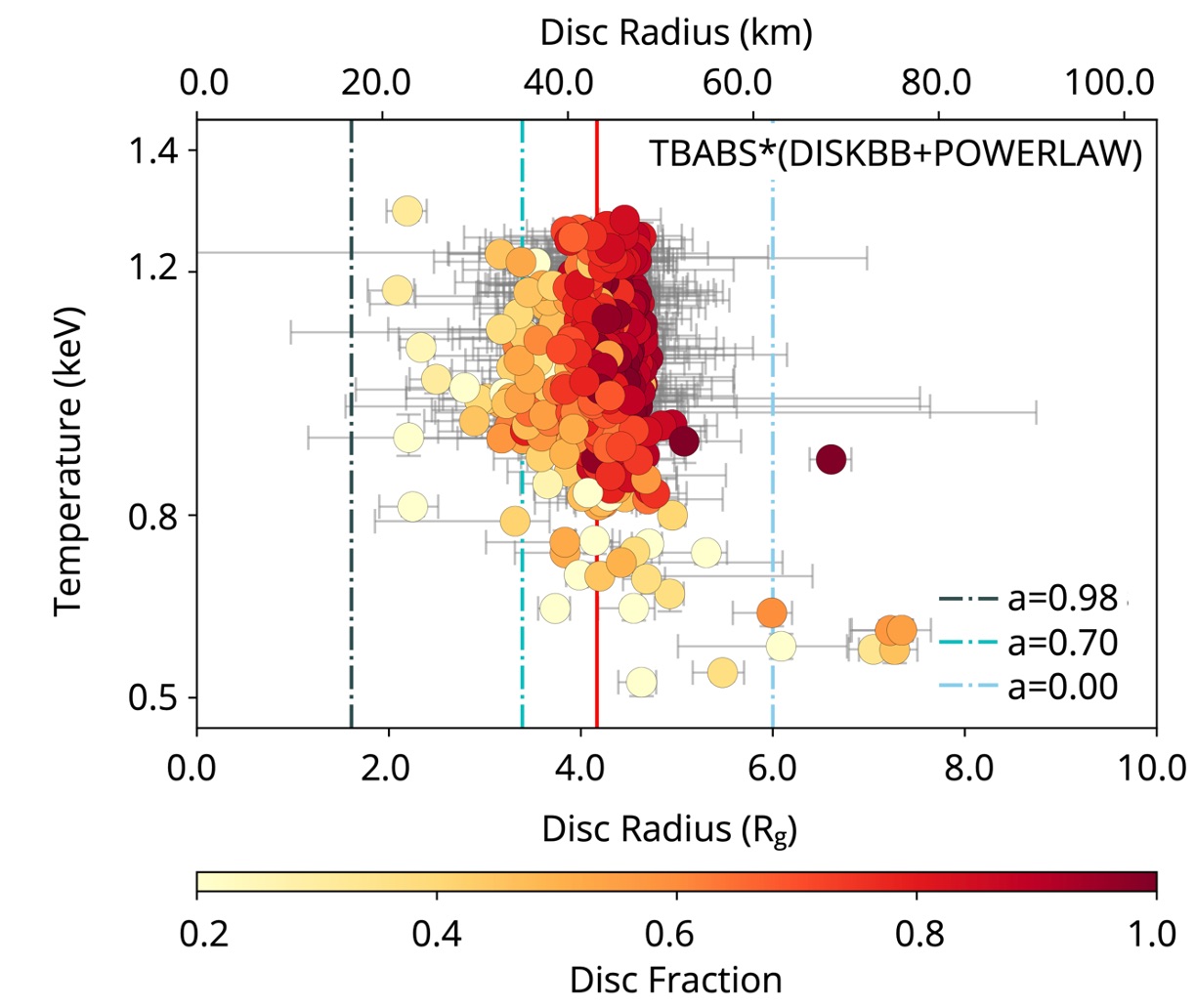}
\includegraphics[width=0.49\textwidth]{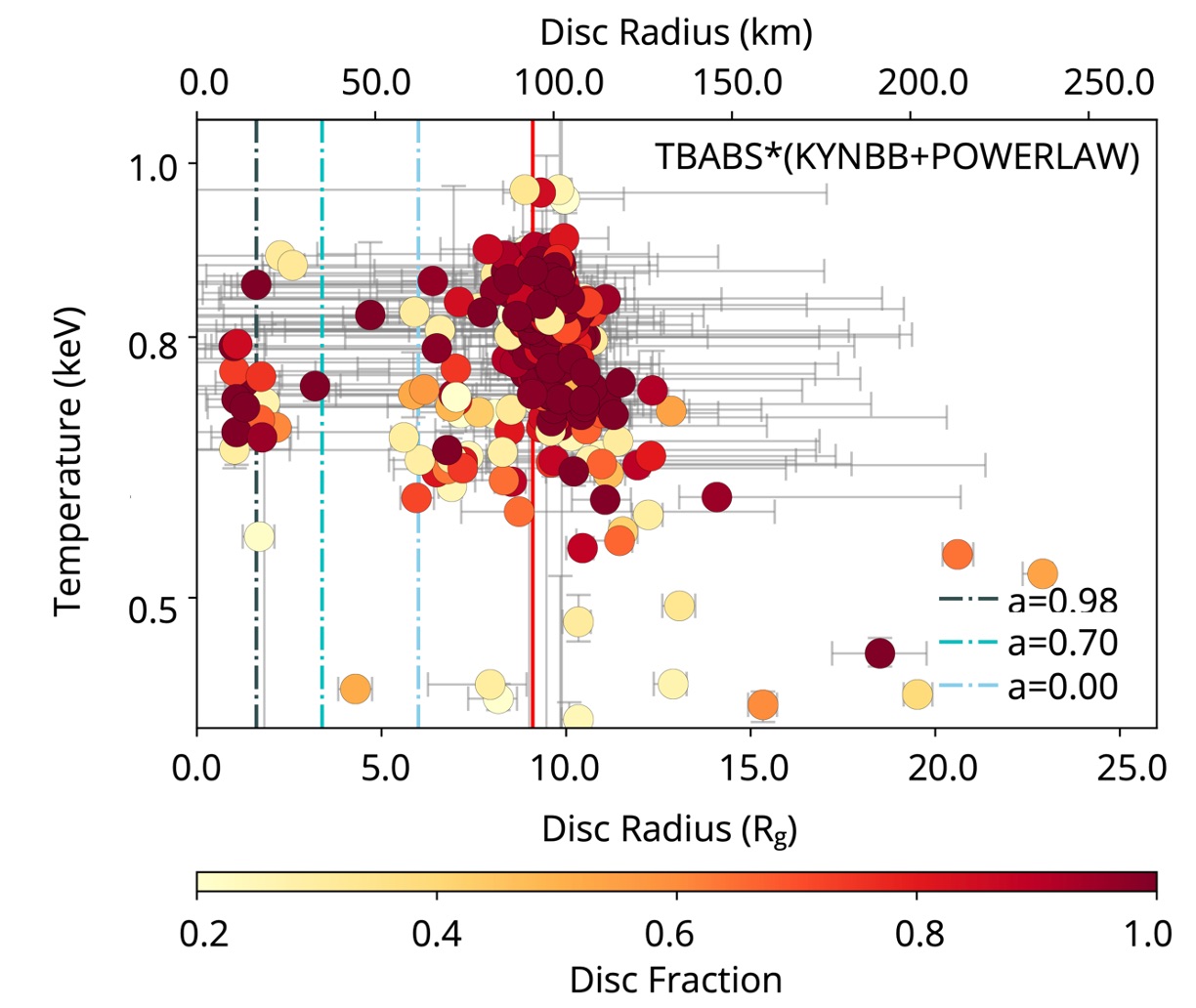}
\caption{\textit{Top}: Change in the inner disc radii with respect to disc temperature for LMC X-3 as measured by \texttt{DISKBB} in units of km \textit{Bottom}: Change in the inner disc radii with respect to disc temperature for LMC X-3 as measured by \texttt{KYNBB} in units of $\rm R_{g}$. Red vertical solid lines indicate the average of the measured inner disc radius values.}
    \label{fig:r_in_lmcx3}
    \textit{} 
\end{figure}

Since \texttt{KERRBB} assumes that $R_{\mathrm{in}}$ always extends down to ${R_{\rm {ISCO}}}$, we compared the ${R_{\rm {ISCO}}}$ values obtained from spin measurements and obtained values 50\% higher on average than ${R_{\rm {in}}}$ values derived from \texttt{DISKBB}. We also investigated the evolution of ${R_{\rm {in}}}$ with respect to ${T_{\rm {in}}}$ throughout all of the outbursts of LMC X-3. LMC X-3 is not observed to follow a similar trend as observed by \texttt{KERRBB} and \texttt{KYNBB} for GRO J1655-40 while similar shifts in values of ${R_{\rm {in}}}$ and ${T_{\rm {in}}}$ can be observed in Figure~\ref{fig:r_in_lmcx3}. 

\subsection{Colour Correction Factor $f_{\mathrm{col}}$}

\begin{figure*}
\centering
\includegraphics[width=0.9\textwidth]{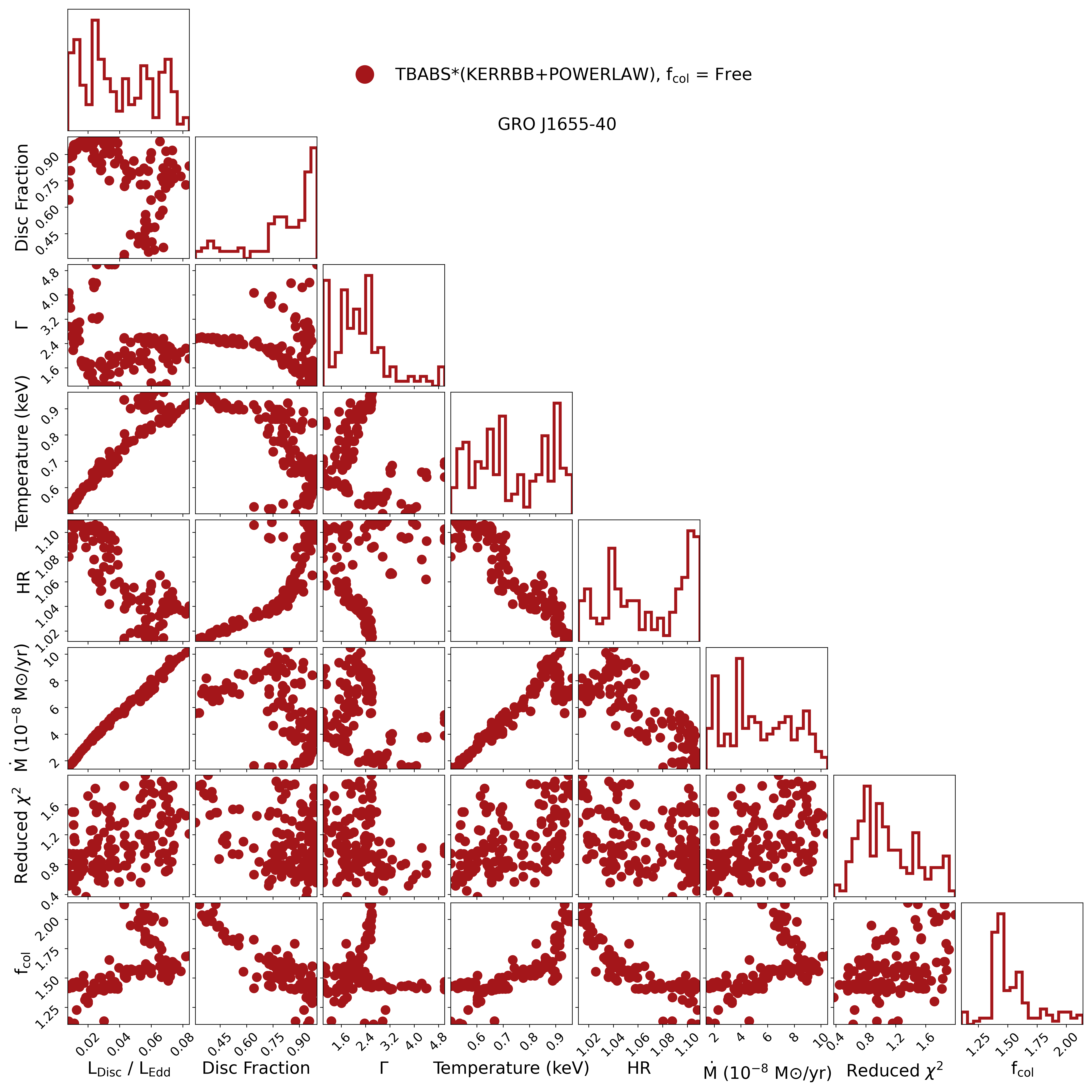}
    \caption{Corner plot showing the evolution of the parameters from our analysis with $f_{\mathrm{col}}$ as a free parameter in \texttt{KERRBB} for GRO J1655-40.}
    \label{fig:corner-f_col-spin}
    \textit{} 
\end{figure*}

Using \texttt{KERRBB}, we investigated the evolution of $f_{\mathrm{col}}$ throughout the 2005 outburst of GRO J1655-40, allowing it to respond to changes in the accretion disc temperature with changing accretion states, following the discussion by \citet{Salvesen13} on using $f_{\mathrm{col}}$ as an alternative to the disc truncation scenario during the state transitions in black hole X-ray binaries. We fixed the spin at 0.7 during this analysis but did not observe the same improvement to the $L_{\rm Disc}$ - $T_{\rm Disc}$ trend to match the reduced $\chi^{2}$ selection criteria as we observed in the previous analysis with spin as a free parameter, in an analogy to changing inner disc radius as discussed in Section~\ref{radius}. We obtained values for $f_{\mathrm{col}}$ covering a range between 1.09-2.12 throughout the outburst. In contrast to the discussion by \citet{Reynolds13, Salvesen21}, we observed a trend where $f_{\mathrm{col}}$ increases with decreasing hardness ratio suggesting a positive correlation between changes in accretion state throughout the outburst and the need for a changing correction factor. We also observed an increase in $f_{\mathrm{col}}$ with an increasing accretion rate and $L_{\mathrm{Disc}}/L_{\mathrm{Edd}}$ with a turnover at $f_{\mathrm{col}} \sim 1.6$ where the accretion rate and $L_{\mathrm{Disc}}/L_{\mathrm{Edd}}$ start to decrease until $L_{\mathrm{Disc}}/L_{\mathrm{Edd}}\sim0.04$ with increasing $f_{\mathrm{col}}$, finally reaching $f_{\mathrm{col}} = 2.12$. This turnover occurs with the observations corresponding to the "spurs" in the top two panels of Figure~\ref{fig:temp-edd-gro}. Figure~\ref{fig:corner-f_col-spin} shows the corner plot presenting the evolution of the parameters from our analysis with $f_{\mathrm{col}}$ as a free parameter. In this analysis, we were able to have a global spin parameter and a steady $R_{\mathrm{ISCO}}$ throughout the outburst of GRO J1655-40 in a similar way observed with variable ${R_{\rm in}}$ in our previous analysis with \texttt{KYNBB} and observed comparably similar $\chi^{2}/\rm{d.o.f.}$ values (within less than \%1 of difference).

 \subsection{Spin Measurements}

\subsubsection{GRO J1655-40}\label{spin-meas-gro}
Our initial analysis with \texttt{KERRBB} across the entire 2005 outburst was based on individual fits of each observation with the spin as a free parameter. We present the distribution of values of the spin parameter measured by \texttt{KERRBB} for both GRO J1655-40 and LMC X-3 in Figure~\ref{fig:spin-kerrbb}. The parameter space for the spin spanned values between $0.5<a_{*}< 0.9$ with a mean at $a_{*}=0.676$. However, such a wide range of parameter values for the spin is not expected to be observed for a black hole even when the statistical deviations arising from the fitting procedure are taken into account. Considering the dependency of spin measurements using the X-ray continuum fitting method on the modelling of the innermost accretion disc and the accretion state, we selected the part of the spectrum where the outburst is on the rise towards and including the HSS. We calculated the hardness ratios (HR) for the entire 2005 outburst where  
\begin{equation}\label{hardness}
 \rm{Hardness\: Ratio (HR)}=\frac{Flux\rm{(6.0-25.0\: keV)}}{Flux \rm{(3.0-6.0\: keV)}}
\end{equation}
Based on the evolution of HR throughout the outburst, we selected a subset of 56 observations with $\mathrm{HR}<0.8$ and $0.05<L_{\rm Disc} / L_{\rm Edd}<0.09$. 
To eliminate any dependencies on other components in the spectra at higher energies, we limited the energy ranges to 3.0-6.0 keV. 
We fit the subset with a free global spin parameter and left the accretion rate and powerlaw component free to vary throughout the sample. We obtained a spin parameter $a_{*}=0.79 ^{+0.012}_{-0.005}$ with $\chi^{2}/\mathrm{d.o.f.} = 543.28/446$. Using the same subset of observations, we obtain a spin parameter $a_{*}=0.74 \pm 0.002$ with \texttt{KYNBB}. Both values are consistent with the results reported by \citet{Shafee06} ($0.65<a_{*}<0.75$ and $a_{*}\sim0.75$ for RXTE and ASCA observations of the 1997 outburst, respectively). The small difference in the measured value by \texttt{KERRBB} most likely arises from the different set of observations used and free $f_{\mathrm{col}}$ used in their analysis with \texttt{KERRBB} (see Section~\ref{f_col} for more discussion on $f_{\mathrm{col}}$ and spin measurements). Additionally, we emphasise that the actual distribution of the spin parameter obtained from individual fits spans a larger range of values between $0.5<a_{*}< 0.9$ than the errors estimated by the global fitting of multiple observations and a more realistic distribution of the parameter could be seen from Figure~\ref{fig:spin-kerrbb}. To be able to obtain a better description of the parameter space of the measured spin from the observations used for the global fit with Xspec, we utilize the nested sampling Monte Carlo algorithm MLFriends \citep{Buchner14,Buchner19} using the UltraNest\footnote{\url{https://johannesbuchner.github.io/UltraNest/}} package \citep{Buchner21} to explore the intrinsic distribution with asymmetric error bars and resample the data points using these error bars. We then assume a Gaussian distribution and obtain the mean and scatter of the distribution ($a_{*}=0.774 \pm 0.069 $ and $a_{*}=0.752 \pm 0.061 $ for \texttt{KERRBB} and \texttt{KYNBB}, respectively). We, therefore, suggest caution when taking smaller-scale error margins obtained from the global fit into consideration. 

\subsubsection{LMC X-3}\label{spin-meas-lmc}
\begin{figure}
\includegraphics[width=0.48\textwidth]{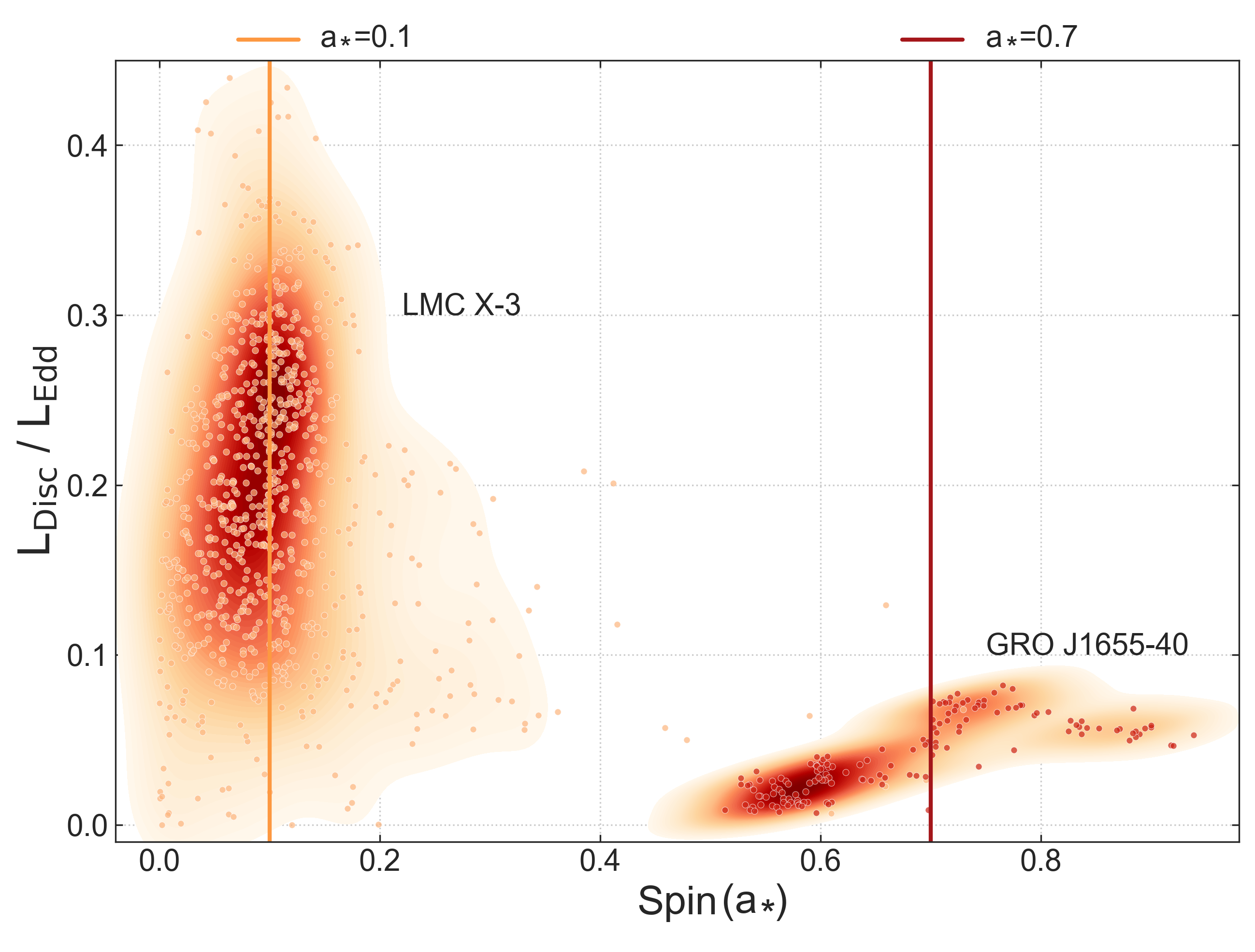}

\includegraphics[width=0.48\textwidth]{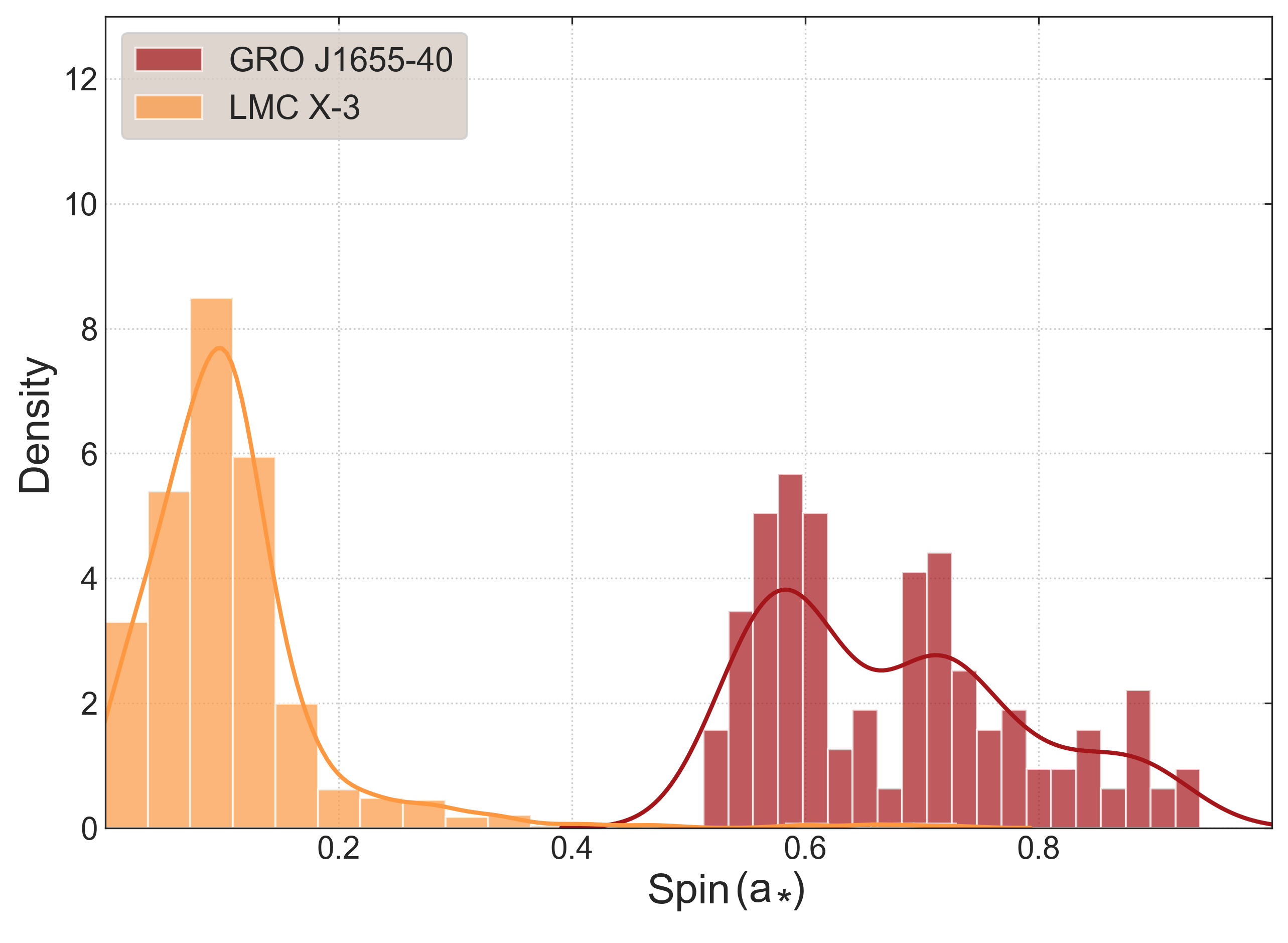}
\caption{\textit{Top}: Evolution of the measured spin values with respect to $L_{\mathrm{Disc}}/L_{\mathrm{Edd}}$ obtained by fitting with \texttt{KERRBB} for LMC X-3 (left) and GRO J1655-40 (right) throughout the entire set of observations listed in Table~\ref{table:obs}.
\textit{Bottom}: Distribution of the black hole spin values as measured by fitting with \texttt{KERRBB} for LMC X-3 (orange) and GRO J1655-40 (red) throughout the entire set of observations described in Section~\ref{spectral_analysis}.}
    \label{fig:spin-kerrbb}
    \textit{} 
\end{figure}

Initial analysis of the entire sample of observations with \texttt{KERRBB} with spin as a free parameter throughout the entire sample of observations revealed dominantly constant spin values with a mean at $a_{*}=0.106$. As previously reported by \citet{Torpin17}, LMC X-3 does not follow the traditional q-shape in the HID as a consequence of its almost "always on" behaviour. They observed different patterns for different cycles as observed by RXTE PCA between 2004-2012. Due to its intrinsically different nature with $L_{\rm Disc} / L_{\rm Edd}$ spanning a much larger parameter space from $\sim$ 0.01 and $\sim$ 0.5 compared to GRO J1655-40 where the system reached $0.082 L_{\rm Disc} / L_{\rm Edd}$ in the highest state, we followed the time evolution of $L_{\rm Disc} / L_{\rm Edd}$ throughout our entire sample (Table~\ref{table:obs}) to define different selection criteria. For simultaneous fitting of LMC X-3, due to the long computational time required for fitting more than 800 spectra with \texttt{KERRBB}, we selected a subset of 72 spectra from one cycle where the observations were dominantly in the soft state with $\mathrm{HR}<0.5$ and $0.1<L_{\rm Disc} / L_{\rm Edd} <0.3$. We restricted the energy range of the fit to 3.0-6.0 keV and measured the spin at $a_{*} = 0.112 ^{+0.003}_{-0.021}$ with $\chi^{2}/\mathrm{d.o.f.} = 296.90/215$. We tested the two different spin cases as reported by \citet{Straub11} and \citet{Steiner14} and compared these fits with our result $a_{*} = 0.112 ^{+0.003}_{-0.021}$. We found that the fit favoured the lower spin case with $a_{*} = 0.1$ over $a_{*} \sim 0.6$ or $a_{*}\sim0.25$ with much lower $\chi^{2}/\mathrm{d.o.f.}$ throughout the entire sample while $a_{*} = 0.1$ is still consistent with the error margin provided by \citet{Steiner14} ($a_{*} = 0.25^{0.20}_{0.29}$). We also analysed the same set of observations with \texttt{KYNBB} and obtained consistent results while leaving the inner disc radius as a free parameter introduced a larger uncertainty on the measured black hole spin which lead to a value of $a_{*} = 0.6 ^{+0.004}_{-0.005} $, in contrast to the low spin case observed by \texttt{KERRBB} and previously reported by \citet{Steiner14}. Similar to spin measurements of GRO J1655-40, these errors estimated on the black hole spin are considerably small compared to the distribution of measured spin values from individual fits and a more realistic distribution of the parameter could be seen from Figure~\ref{fig:spin-kerrbb}. We employed the same approach as described in Section~\ref{spin-meas-gro} and obtained $a_{*}\sim0.098 \pm 0.063$ for \texttt{KERRBB}. We again suggest caution when considering significantly smaller-scale error margins obtained from global fitting with Xspec. 

\citet{Yilmaz23} provides an extensive analysis of the 2005 outburst of GRO J1655-40 using the same observations presented in Table~\ref{table:obs} to compare \texttt{KERRBB} and \texttt{KYNBB} when the innermost edge in \texttt{KYNBB} is set at ISCO and showed that both models are identical and the minor differences in the measured parameter values between \texttt{KERRBB} and \texttt{KYNBB} can be explained by small statistical fluctuations arising from the fitting procedure with Xspec. While GRO J1655-40 provided an ideal set of observations to study the differences as a result of changing boundary conditions when the inner edge of the disc is no longer at ISCO, this significant difference in the measured black hole spin by \texttt{KERRBB} and \texttt{KYNBB} with $R_{in}>R_{ISCO}$ can be explained by the larger error margins in the measured inner disc radii introduced by the reduced signal-to-noise ratio of LMC X-3 spectra due to its distance when combined with the instrumental uncertainties. It is, therefore, to be taken into consideration with caution.

\begin{figure}
\includegraphics[width=0.48\textwidth]{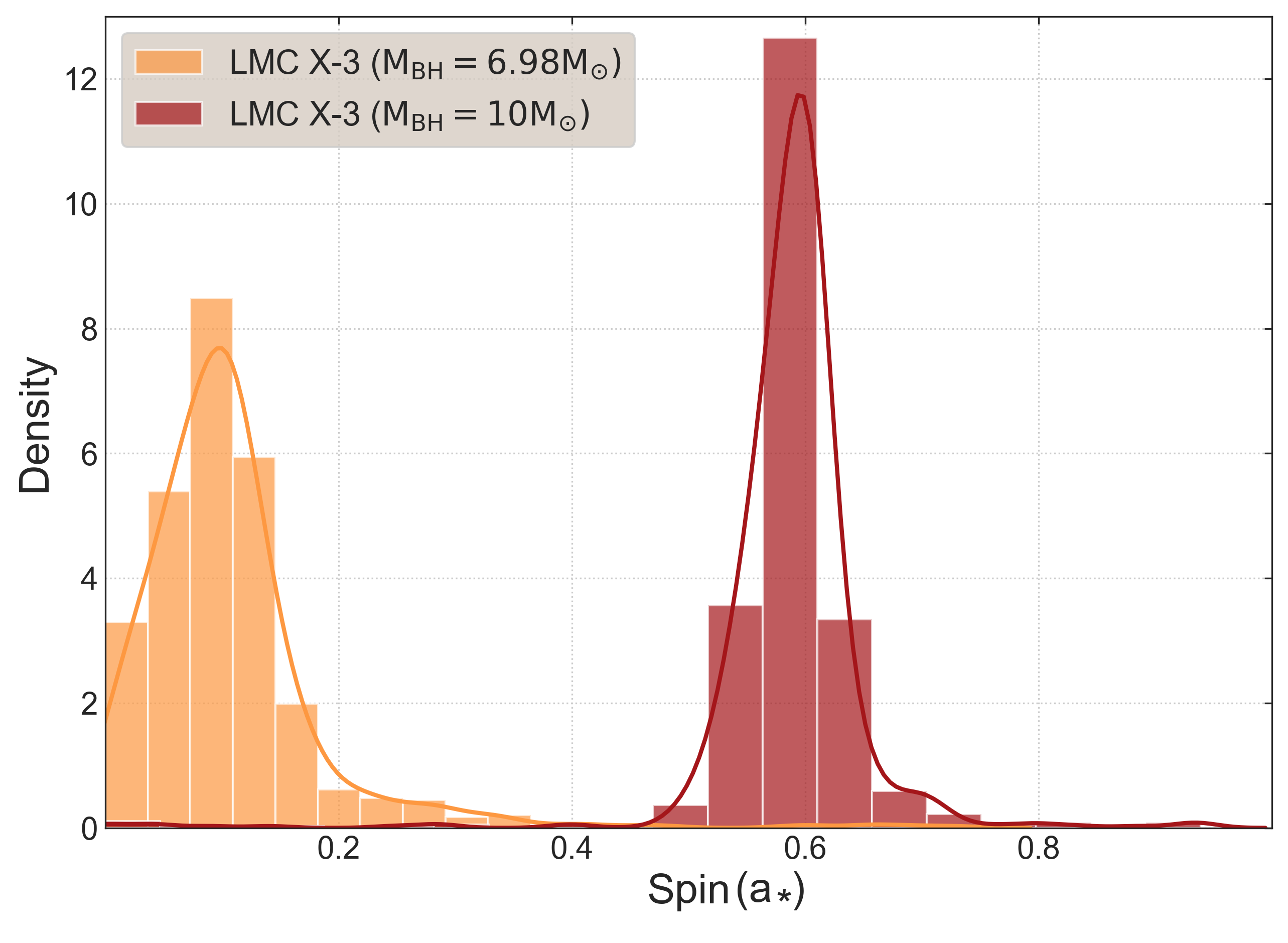}
\caption{Distribution of black hole spin values as measured by \texttt{TBABS$\times$(KERRBB+POWERLAW)} throughout the entire set of observations of LMC X-3 for $M_{\mathrm{BH}}=6.98 M_{\odot}$ (orange) and $M_{\mathrm{BH}}=10 M_{\odot}$ (red).}
    \label{fig:spin-comparison}
    \textit{} 
\end{figure}

This contrast in the measured spin by \texttt{KERRBB} (when compared with \citet{Straub11}) is dominantly due to the difference in the assumed black hole mass and inclination angle in the analysis. Before a more constrained update was provided by \citet{Orosz14}, the study of the inner disc radius and spin measurements by \citet{Steiner10} and \citet{Straub11}, respectively, was based on a mass of $M_{\mathrm{BH}}=10 M_{\odot}$. In this study, we adopt the more recent value of $M_{\mathrm{BH}}=6.98 \pm 0.56 M_{\odot}$. We tested the effect of this difference by setting black hole mass in the analysis described above with \texttt{KERRBB} to $M_{\mathrm{BH}}=10 M_{\odot}$ and we obtained $a_{*} = 0.601 ^{+0.011}_{-0.009}$. On the other hand, the effect of inclination angle was not as strong in the measured spin for $M_{\mathrm{BH}}=6.98 M_{\odot}$. Similar to the analysis described previously, we didn't observe the same deviations in the measured spin as previously reported by \citet{Straub11}. Figure~\ref{fig:spin-comparison} shows the difference in the distribution of black hole spin parameter as measured by \texttt{KERRBB} throughout the entire set of observations of LMC X-3 for two black hole masses ($M_{\mathrm{BH}}=6.98 M_{\odot}$ (orange) and $M_{\mathrm{BH}}=10 M_{\odot}$ (red)). 

\section{DISCUSSION}\label{discussion}

We investigated the evolution of the accretion discs throughout multiple outbursts of GRO J1655-40 and LMC X-3 by analysing $\sim$ 1800 spectra obtained using RXTE with a wide range of disc luminosities of $1.0\times10^{-4}<L_{\rm Disc} / L_{\rm Edd}<0.5$.

\texttt{DISKBB} has been used extensively in many spectral studies of accretion discs in BHXRBs and was found to provide a significantly reasonable approximation for the spectral shape of a geometrically thin and optically thick Newtonian \citet{Shakura73} disc. It was criticized by \citet{Gierlinski01} for ignoring the inner torque-free boundary condition at the inner edge of the accretion disc. This non-zero torque assumption at the inner boundary of the accretion disc was put into test by \citet{Zimmerman05} where they presented a comparison of \texttt{DISKBB} with its zero-torque counterpart \texttt{EZDISKBB}. They showed that \texttt{DISKBB} temperature ($T_{\mathrm{max}}$) values were $\approx 5\%$ higher on average than those obtained by fitting with \texttt{EZDISKBB}. In contrast, they observed higher values for $R_{\mathrm{in}}$ with \texttt{DISKBB} by more than a factor of $\sim$ 2.17 on average. They showed that the application of a zero-torque boundary condition has little to no effect on an accretion disc temperature peaking at $\sim$ 1.8 keV as observed in GRO J1655-40 with \texttt{DISKBB} while the biggest impact is on the inner accretion disc radius.

In comparison to the relativistic disc models, \texttt{DISKBB} was shown to produce significantly higher disc temperatures for two main reasons:
\begin{enumerate}
 \item It assumes a perfect blackbody emission from the accretion disc independent of the accretion state the system is in, not accounting for non-blackbody emission arising from the complicated scatterings in the accretion disc atmosphere. This assumption essentially means $f_{\mathrm{col}} = 1.0 $ for the fit while in $f_{\mathrm{col}} \sim 1.7$ is widely accepted to provide a correction to deviations of the local accretion disc spectrum from the "perfect blackbody" emission for the unaccounted scattering and Comptonisation processes in the disc atmosphere. 
 \item \texttt{DISKBB} provides a relatively simplistic description of a multi-colour accretion disc without any general relativistic effects accounting for strong gravitational fields around black holes. These relativistic effects include frame-dragging, returning radiation, Doppler boost, gravitational redshift, self-irradiation etc. which are needed to provide a better solution for the emitted spectrum from the innermost region of the accretion disc around a black hole of any mass. These corrections modify the calculated local specific intensity as the photons escape the gravitational potential of the black hole and eventually reach the observer. Before any relativistic corrections to \texttt{DISKBB} were made available, some attempts have been made to adopt the correction by \citet{Zhang97} to the observed temperature while \citet{Li05} compared the results of such applications with four disc models and found that the approach was still lacking in the treatment of relativistic processes on the calculated spectra.
\end{enumerate}

Our investigation showed that a simple change from the non-relativistic \texttt{DISKBB} model to \texttt{KERRBB} did not improve the trend except for a significant decrease in the disc temperatures throughout the outburst. This trend was significantly improved by employing \texttt{KYNBB} with $R_{\mathrm{in}}$ allowed to vary with the state transitions. This behaviour suggests a dominating power of $R_{\mathrm{in}}$ in combination with general relativistic effects for the states with the smallest measured inner disc radii. For these states, \texttt{KERRBB} measured spin values as high as $0.8 < a_{*} < 0.98$ which correspond to a region $1.6\: R_{\rm g} < R_{\rm ISCO} < 2.9\: R_{\rm g}$. 

RXTE stood out as one of the most successful X-ray missions in the study of stellar-mass black holes in X-ray binaries but possesses a limiting factor due to its limited energy resolution and poor effective area at lower energies where the thermal emission from the accretion disc is expected to peak. It has an advantageous position in the field with its ability to observe a significant number of sources throughout many outbursts making it an important tool for the purpose of studying the evolution of accretion states in black hole X-ray binaries. RXTE PCA is sensitive to energy ranges above $\sim$ 3 keV. This still corresponds to a range where the tail of the thermal emission can be modelled with \texttt{DISKBB} as the source transitions to the HSS throughout the outburst while it can be extended to around 6-7 keV with its relativistic counterparts \texttt{KERRBB} and \texttt{KYNBB}. To test the effects of these instrumental limitations, we made use of 6 out of 39 publicly available Swift observations of GRO J1655-40 which could be modelled by the same model setups adopted throughout our analysis and obtained consistent results in the parameter values presented in this work. Additional to Swift observations, we made use of INTEGRAL/ISGRI which is much better calibrated at higher energies to test the effect of the contribution of high energy emission on measured parameter values. We used 6 sets of observations obtained simultaneously to compare the measured parameter values of disc temperatures, disc radii and spin using \texttt{DISKBB} and \texttt{KERRBB} and obtained consistent values. For simplicity, we exclude these observations from further discussion.

X-ray continuum fitting method to measure the spin depends on determining the inner edge of the accretion disc $R_{\rm in}$ based on the assumption that the disc always extends down to the $R_{\rm ISCO}$ in the HSS and inferring the spin using  equation~(\ref{r_isco}). However, this method has its limitations even after careful treatment of $f_{\mathrm{col}}$. Even though the most commonly accepted value of $f_{\mathrm{col}}$ stands at 1.7, the precise determination of this correction factor for each observed accretion state still prevails to be beyond the bounds of what current instruments and methods allow. Both \texttt{KERRBB} and \texttt{KYNBB} depend on how precisely the mass, inclination angle and distance to the source are measured. Our analysis of GRO J1655-40 with \texttt{KYNBB} showed a wide range of inner disc radii. While these radii are observed to be much larger than $R_{\mathrm{ISCO}}$ for $0.8 < a_{*} < 1.0$, it's worth keeping in mind that these observations were obtained using RXTE PCA spectra limited to 3.0-25.0 keV which delivers extra uncertainties to these measurements. The much larger scatter in $R_{\mathrm{in}}$ vs Temperature can be interpreted as these uncertainties since measuring the inner disc radius with much lower temperatures than previously observed will be compromised by PCA's limited energy range.

\subsection{Comparing GRO J1655-40 and LMC X-3} 

Both GRO J1655-40 and LMC X-3 are ideal candidates to study the evolution of accretion discs around black holes in XRBs. Both are known to harbour a black hole of similar mass and have similar inclination angles (Table~\ref{table:src}). LMC X-3 stood out as a perfect source for black hole spin measurements due to its dominantly stable inner disc radius. Our analysis of LMC X-3's inner disc radius with \texttt{DISKBB} and \texttt{KYNBB} yielded consistent results to previous studies (Figure~\ref{fig:r_in_lmcx3}). Despite this large difference in $\dot{M}$, the innermost edge of the accretion disc doesn't show as much variance as observed from GRO J1655-40. LMC X-3 also doesn't show the reported "spurs" in $L_{\rm Disc}$ - $T_{\rm Disc}$ in contrast to GRO J1655-40 and other BHXRBs reported by \citet{Dunn11}. Considering the changes to this relationship introduced by relativistic modelling of the accretion disc using \texttt{KYNBB} with free $R_{\rm in}$ (Figure~\ref{fig:temp-edd-gro}), this can indicate either a strong dependency on the inner disc radius evolution or weaker relativistic effects acting on the observed spectrum due to the low spin of LMC X-3 and the accretion disc being truncated at larger radii (Figure~\ref{fig:r_in_lmcx3}).

Each corner plot in Figure~\ref{fig:all-both} shows the evolution of relevant parameters in our analysis with \texttt{TBABS$\times$(DISKBB+POWERLAW)}, \texttt{TBABS$\times$(KERRBB+POWERLAW)} and \texttt{TBABS$\times$(KYNBB+POWERLAW)} for GRO J1655-40 (red) and LMC X-3 (orange), respectively. Independent of the nature of the disc model, RXTE observations of LMC X-3 have shown that the source evolves through a larger parameter range for both $L_{\rm Disc}$ and $\dot{M}$ compared to GRO J1655-40. This larger parameter range translates to a difference in $\dot{M}$ by about an order of magnitude for both \texttt{KERRBB} and \texttt{KYNBB}. Both GRO J1655-40 and LMC X-3 were observed to show changes in the inner disc temperature $T_{\rm Disc}$ throughout the entire set of observations, covering multiple accretion states. While this change in $T_{\rm Disc}$ was directly reflected in the evolution of $R_{\rm in}$ in GRO J1655-40, a similar evolution was not observed for LMC X-3 with both \texttt{DISKBB} and \texttt{KYNBB}. There's a clear trend between the disc fraction and $T_{\rm Disc}$ for GRO J1655-40 however, it's not easy to observe a clear trend in disc fraction with respect to any other parameters for all of the models described above. We also acknowledge the unexpected inverse relationship between the disc fractions and measured $R_{\rm in}$. According to the standard scenario describing the evolution of thermal and non-thermal components, smaller disc radii are expected to correspond to more disc-dominated states while a disc truncated at a larger distance is usually associated with a powerlaw-dominated state. Our analysis showed a direct relationship between the hardness ratio and disc fractions where increasing disc fractions correspond to higher hardness ratios. While we observed a similar during the entirety of our analysis, we present it only in Figure~\ref{fig:corner-f_col-spin} and exclude this evolution for each model to avoid increasing the complexity of the parameter space presented in each corner plot. This trend is not naturally expected and might be explained by a powerlaw component dominating the disc emission when extrapolated to lower energies for a specific photon index while both the spectral shapes and hardness ratios point towards softer spectra and hence lower hardness ratios in comparison to higher disc fractions. Instead of using the empirical \texttt{POWERLAW}, it is usually suggested to adopt a Comptonisation model \texttt{SIMPL} \citep{Steiner09} to overcome this divergence of the powerlaw component at lower energies, though a simple interchange of these models produced worse fits for the majority of our sample. Therefore, a detailed investigation of the powerlaw component is out of the scope of the analysis presented here.

Regardless of the contrasting measurements of the black hole spin reported over the past decade for LMC X-3 (see discussion in Section~\ref{sources}), our results show a polarized distribution of the spin parameter across observations (Figure~\ref{fig:spin-kerrbb}). The distribution of the measured spin values of LMC X-3 is more concentrated at around $\sim$ 0.1 while GRO J1655-40 shows a bimodal distribution of the spin parameter throughout the entire outburst. This bi-modality in measured spin values of GRO J1655-40 corresponds to the two sets of observations that can be grouped by whether these observations are on the expected $L_{\rm Disc}$ - $T_{\rm Disc}$ line or the so-called "spurs". The higher spin case ($a_{*}\sim$0.8) dominantly corresponds to these "spurs" while the lower end of the distribution ($a_{*}\sim$0.6) corresponds to the rest of the observations. A more detailed evolution of the spin parameter can be seen from the top plot of Figure~\ref{fig:r_in-time-gro}. These spurs are also the observations where the inner edge of the disc is observed to be the closest to the black hole, which can explain the observed high spin by \texttt{KERRBB} while disc fractions are significantly lower when compared to the rest of the observations. The absence of such clear evolution in $R_{\rm in}$ in LMC X-3 is observed in a more or less steady black hole spin throughout the entire set of observations covering multiple outbursts. This also could explain the absence of spurs in LMC X-3. From our analysis, it's clear that these spurs are not luminosity dependent and a simple increase in the mass accretion rate or disc luminosity cannot explain their presence in the majority of 25 BHXRBs as presented in \citet{Dunn11}. The distribution of measured inner disc radii for GRO J1655-40 showed a significant variation in the parameter for similar disc fractions, suggesting the possibility of evolving inner disc even within an accretion state. Our analysis also revealed that the disc might not be extending all the way down the ISCO even in the highest luminosity states. As can be observed from Figure~\ref{fig:r_in-temp_gro}, the lowest inner disc radius in GRO J1655-40 measured by \texttt{KYNBB} is a few gravitational radii away from ISCO for both $a_{*}\sim$0.7 and $a_{*}\sim$0.8. While constraining $R_{\rm in}$ in LMC X-3 with \texttt{KYNBB} was rather difficult, a similar trend was observed in which a significant portion of the values correspond to $R_{\rm in} > R_{\rm ISCO}$ for $a_{*}\sim$0.1. This peculiar behaviour of $R_{\rm in}$ introduces strong implications for the reliability of the X-ray spectral continuum method for black hole spin measurements.

A more careful investigation of reduced $\chi^{2}$ distribution  revealed no clear evidence for a bias arising from the selection criteria adopted with a cut at reduced $\chi^{2} = 2$  (Figure~\ref{fig:all-both}). While such investigation could prove to be useful when assessing the presence of emission lines and the appropriate modelling of such features, specific observations with strong Fe emission lines (especially during the 1996-1997 outburst of GRO J1655-40) were excluded to avoid introducing the risk of any selection bias and model dependency.

\subsection{Relativistic vs. Non-Relativistic $R_{\rm in}$ Measurements}

Expectedly, the inner disc radius measurements of GRO J1655-40 from spectral fitting with the non-relativistic disc model \texttt{DISKBB} produced results ranging between $\sim$10 and $\sim$30 km throughout the 2005 outburst, much smaller than $R_{\rm ISCO}$ values for $0.0<a_{*}<0.85$ which is the range where all reported spin values obtained from X-ray continuum method lie. With significantly higher disc temperatures, $\sim$10\% of the observations yields inner disc radii corresponding to a region as close as $\sim$1-2 $R_{\rm g}$. This region corresponds to a region within $R_{\rm ISCO}$ for a moderately spinning black hole where the matter around the black hole can no longer sustain orbital motion and starts spiralling down to the black hole. While the outer radius of the disc can extend out to hundreds and even thousands of $R_{\rm g}$, inner disc radii residing within this region for the entirety of the outburst covering disc fractions from 0.1 to 1.0 does not produce a physically reasonable picture for the accretion disc geometry even for significantly higher spin parameters. Such small radii from the normalisation of \texttt{DISKBB} result from the non-relativistic treatment of the disc and not accounting for the non-blackbody emission due to scattering and Comptonisation processes in the accretion disc atmosphere. Changes in disc fractions throughout all of the outbursts demonstrated that even a temperature difference of $\sim$ 0.2 keV becomes significant to increase the disc fraction from $\sim$0.3 to $\sim$0.6 switching from a powerlaw dominated spectrum to a moderately disc-dominated spectrum. Calculated throughout the same outbursts, this difference corresponds to a 50\% higher temperatures obtained from \texttt{DISKBB} and this results in $\sim$5 times larger inner disc radii.

There have been attempts to include the necessary relativistic corrections following \citet{Zhang97,Kubota98, Gier04} while the main approach has been focused on a correction applied post-analysis using a constant calculated for a set of certain parameters (See \citet{Li05} and references therein). While such corrections provide improvements to a certain degree, the post-analysis approach is lacking in the proper application of relativistic corrections as it is based on adopting a factor calculated for a set of parameter values and does not take into account the fact that any relativistic correction should have a strong dependency on the distance to the black hole and hence should correspond to different values for each observation throughout an outburst for sources that are known to exhibit a great evolution of the inner disc radius like GRO J1655-40. On the other hand, taking the example of \citet{Kubota98}, the calculated correction factor values for a set of $f_{\mathrm{col}}$ still remain not enough to shift the low values of radii measured by \texttt{DISKBB} or similar non-relativistic disc models.

As can be seen from the bottom plot of Figure~\ref{fig:r_in-temp_gro}, the needed correction is much larger than what was calculated by \citet{Zhang97,Kubota98, Gier04}. These results suggest that the need for relativistic corrections are much greater than what a simple factor can provide and needs to be handled by self-consistent disc models such as \texttt{KERRBB} and \texttt{KYNBB} which adopts the ray tracing method with radius-dependent corrections taken into account while calculating the spectrum.

\subsection{Effects of $f_{\mathrm{col}}$ on the Disc Temperature}\label{f_col}

Apart from the relativistic treatment provided by \texttt{KERRBB} and \texttt{KYNBB} to the observed spectra from the accretion discs around black holes, $f_{\mathrm{col}}$ is the second important parameter that sets relativistic and \texttt{DISKBB} apart. This key factor provides the correction to the spectrum to account for electron scatterings in the accretion disc atmosphere causing the spectrum to deviate from the perfect blackbody emission. The need for such a parameter arises where the gas temperature at the surface of the disc is high enough such that the electron scatterings start to dominate over absorptive processes taking place in the disc atmosphere. This corresponds to photon energies > 1.0 keV \citep{Shakura73}. Both \texttt{KERRBB} and \texttt{KYNBB} calculate the observed spectrum with $f_{\mathrm{col}}$ applied and has this correction factor as a model parameter. Determining a constrained value of the $f_{\mathrm{col}}$ through spectral modelling using accretion disc models like \texttt{KERRBB} and \texttt{KYNBB} is greatly affected by the strong degeneracy between the $f_{\mathrm{col}}$ and the black hole spin \citep{Nowak08, Nowak12}. \citet{Salvesen21} presented the level of sensitivity of black hole spin measurements using the X-ray spectral continuum method, suggesting a direct relationship between $f_{\mathrm{col}}$ and the measured black hole spin. As a result of this degeneracy, it requires extra caution during any attempt to determine the true value of $f_{\mathrm{col}}$ for a specific observation in a specific accretion state.

Figure~\ref{fig:corner-f_col-spin} shows the relationship between $f_{\mathrm{col}}$ and other parameters. We found values of $f_{\mathrm{col}}$ spanning the range 1.09-2.12, slightly broader than previously reported 1995 and 1997 ASCA and 1997 RXTE observations of GRO J1655-40 by \citet{Shafee06} and calculated best-fit values from \texttt{BHSPEC} \citep{Davis05}(1.3-1.9). Our findings in terms of hardness ratio vs. $f_{\mathrm{col}}$ contradict the suggested scenario where $f_{\mathrm{col}}$ is expected to decrease as the source transitions to softer states to account for a non-truncated disc radius in the LHS \citep{Reynolds13}. In contrast, we observed a decrease in $f_{\mathrm{col}}$ with hardening spectra.

Our analysis showed that adopting the canonical value for $f_{\mathrm{col}} (T_{\mathrm{col}} / T_{\mathrm{eff}})$ at 1.7 following \citet{Shimura95} cannot explain a difference of only 50\% in disc temperatures between \texttt{DISKBB} and \texttt{KERRBB} and/or \texttt{KYNBB} on its own. The top panel of Figure~\ref{fig:temp-edd-gro} shows the observed $L_{\rm Disc}$ - $T_{\rm Disc}$ relationship for GRO J1655-40 and dashed lines represent $L_{\rm Disc}/L_{\rm Edd}$ for different $f_{\mathrm{col}}$ calculated using  equation~(\ref{eddington_ratio}). $L_{\rm Disc}/L_{\rm Edd}$ - $T_{\rm Disc}$ relationships obtained from both relativistic and non-relativistic disc models corresponded to larger values of $f_{\mathrm{col}}$ independent of its set values in \texttt{KERRBB} and \texttt{KYNBB}. While our analysis with variable $R_{\rm in}$ in \texttt{KYNBB} and variable $f_{\mathrm{col}}$ in \texttt{KERRBB} produced indistinguishable results in reduced $\chi^{2}/\rm{d.o.f.}$ values, the latter model setup did not provide an improvement to the previously discussed deviations in Figure~\ref{fig:temp-edd-gro}.

\section{SUMMARY}\label{summary}

In this paper, we performed a comprehensive spectral analysis of the publicly available RXTE observations of GRO J1655-40 and LMC X-3 between 1996 and 2011, investigating the behaviour of the accretion disc temperature and radius by comparing relativistic and non-relativistic disc models \texttt{DISKBB}, \texttt{KERRBB} and \texttt{KYNBB} with the following conclusions:

\begin{enumerate}

\item To test the possible contribution to the emitted spectrum by the general relativistic effects resulting from the strong gravitational potentials in the close vicinity of black holes, we investigated the same RXTE observations of GRO J1655-40 using \texttt{KERRBB} and \texttt{KYNBB}. Simply adopting \texttt{KERRBB} instead of \texttt{DISKBB} did not provide any improvements to the deviations in $L_{\rm Disc}$ - $T_{\rm Disc}$ relationship (Section~\ref{t_lumin}). These deviations, however, were not detected when $R_{\rm in}$ was defined as a free parameter in \texttt{KYNBB}. 

\item The inner disc radii calculated from the normalisation parameter of the \texttt{DISKBB} (see  equation~(\ref{normalization-diskbb})) are systematically lower than the values obtained from the relativistic disc models. \texttt{KYNBB} measured much larger inner disc radii throughout the entire outburst of GRO J1655-40 by at least a factor of 6 on average. This difference was even larger for LMC X-3, a factor of $\sim$8 while the uncertainties in these measurements were much larger compared to GRO J1655-40 (Section~\ref{radius}). These larger uncertainties observed for LMC X-3 can be due to degeneracies between model parameters and poorer quality of spectra with much lower flux values by a factor of 100-1000 (compared to the spectra from the 2005 outburst of GRO J1655-40), on average throughout different accretion states.

\item The colour correction factor $f_{\mathrm{col}}$ (with the spin parameter fixed) is found to evolve throughout an outburst spanning a range between 1.0 and 2.2. This evolution follows a pattern where $f_{\mathrm{col}}$ increases with increasing accretion rate/Eddington ratio before reaching a turning point at $f_{\mathrm{col}}\sim1.6$ and accretion rate/Eddington ratio starts to decrease with $f_{\mathrm{col}}$ still increasing (Section~\ref{f_col}). Our analysis showed that both $f_{\mathrm{col}}$ and $R_{\rm in}$ as free parameters performed equally acceptable when the reduced $\chi^{2}/\rm{d.o.f.}$ values from each analysis are compared while adopting a variable $f_{\mathrm{col}}$ throughout the outburst did not provide the same improvement to the deviations in $L_{\rm Disc}$ - $T_{\rm Disc}$ observed in \texttt{DISKBB} and \texttt{KERRBB}. This suggests strong support for a scenario where the changing inner disc radius coupled with general relativistic effects to explain these deviations observed in GRO J1655-40.

\end{enumerate}

\section*{ACKNOWLEDGMENTS}

The authors thank Enrico Bozzo and Carlo Ferrigno for technical support with the analysis of INTEGRAL data and fruitful suggestions and discussions. The research leading to these results has received funding from the European Union’s Horizon 2020 Programme under the AHEAD2020 project (grant agreement n. 871158) based on observations with INTEGRAL, an ESA project with instruments and science data centre funded by ESA member states (especially the PI countries: Denmark, France, Germany, Italy, Switzerland, Spain), and with the participation of the Russian Federation and the USA. A.Y. acknowledges the support from GAUK project No. 102323. A.Y. and P.G.B. acknowledge financial support from the Czech Science Foundation under Project No. 19-05599Y. M.B., M.D. and J.S. acknowledge the support from the GACR project 21-06825X. A.Y., P.G.B., M.B., M.D., and J.S. also acknowledge the institutional support from RVO:6798581. The authors also would like to thank the anonymous referee for the valuable and detailed comments and suggestions.
\section*{DATA AVAILABILITY}
All of the RXTE data used in this paper (see Table~\ref{table:obs}) are publicly available in HEASARC's archive (\url{https://heasarc.gsfc.nasa.gov/docs/archive.html}).




\bibliographystyle{mnras}
\bibliography{main} 




\appendix
\section{Appendix}

\begin{figure}
\centering
\includegraphics[width=0.48\textwidth]{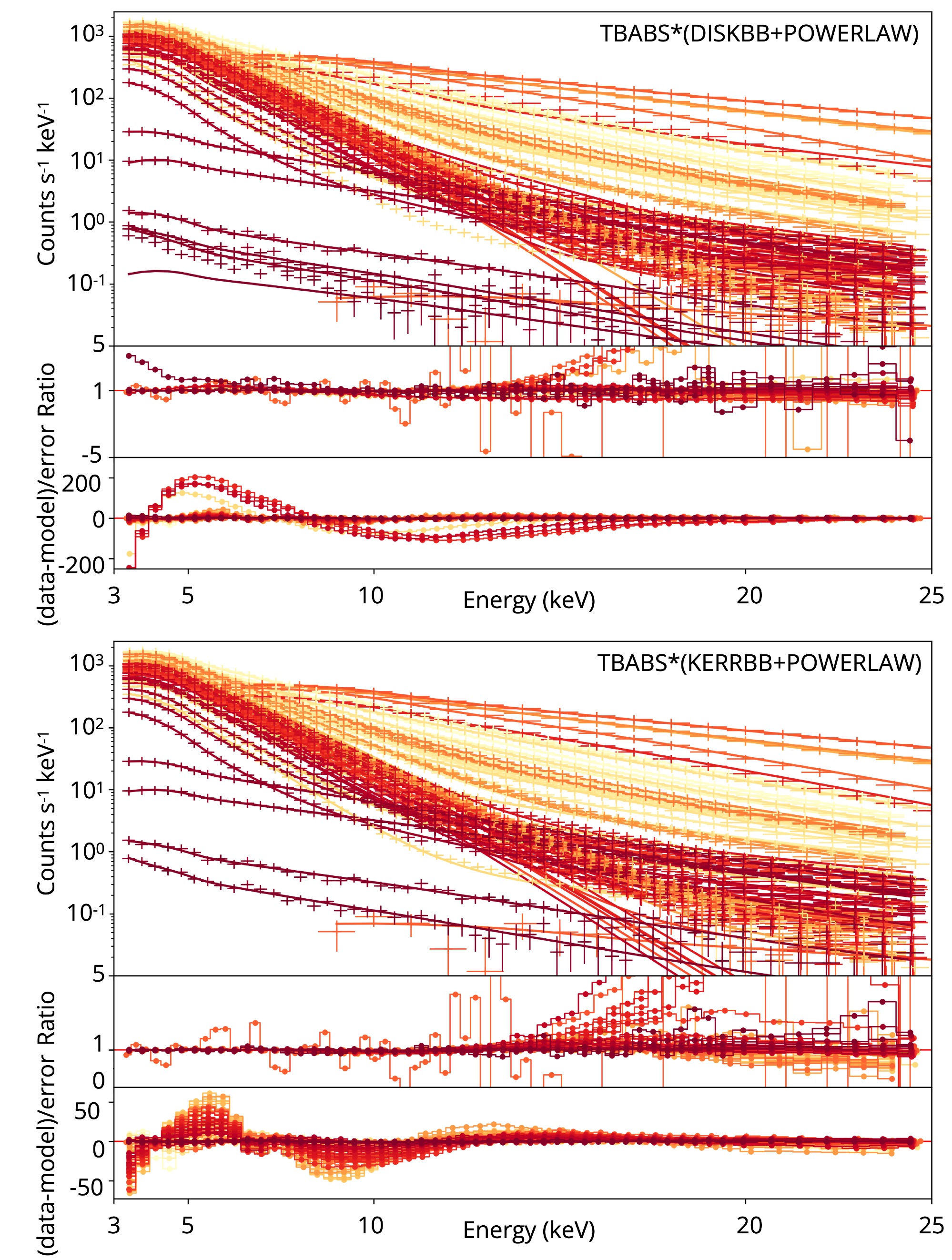}
\caption{\textit{Top}: Data and total model obtained from spectral fitting of GRO J1655-40 during 1996 and 1997 outbursts that produced significantly worse fits with reduced $\chi^{2}\gg2.0$ obtained from spectral fitting with \texttt{TBABS$\times$(DISKBB+POWERLAW)}, showing the residuals in ratios (data/model) and (data-model)/error below.
\textit{Bottom}: Data and total model obtained from spectral fitting of GRO J1655-40 during 1996 and 1997 outbursts that produced significantly worse fits with reduced $\chi^{2}\gg2.0$ obtained from spectral fitting with \texttt{TBABS$\times$(KERRBB+POWERLAW)}, showing the residuals in ratios (data/model) and (data-model)/error below.
}
    \label{fig:spec-gro-96-97}
    \textit{} 
\end{figure}

\begin{figure*}
    \begin{subfigure}{1.\textwidth}
        \centering
        \includegraphics[width=\textwidth]{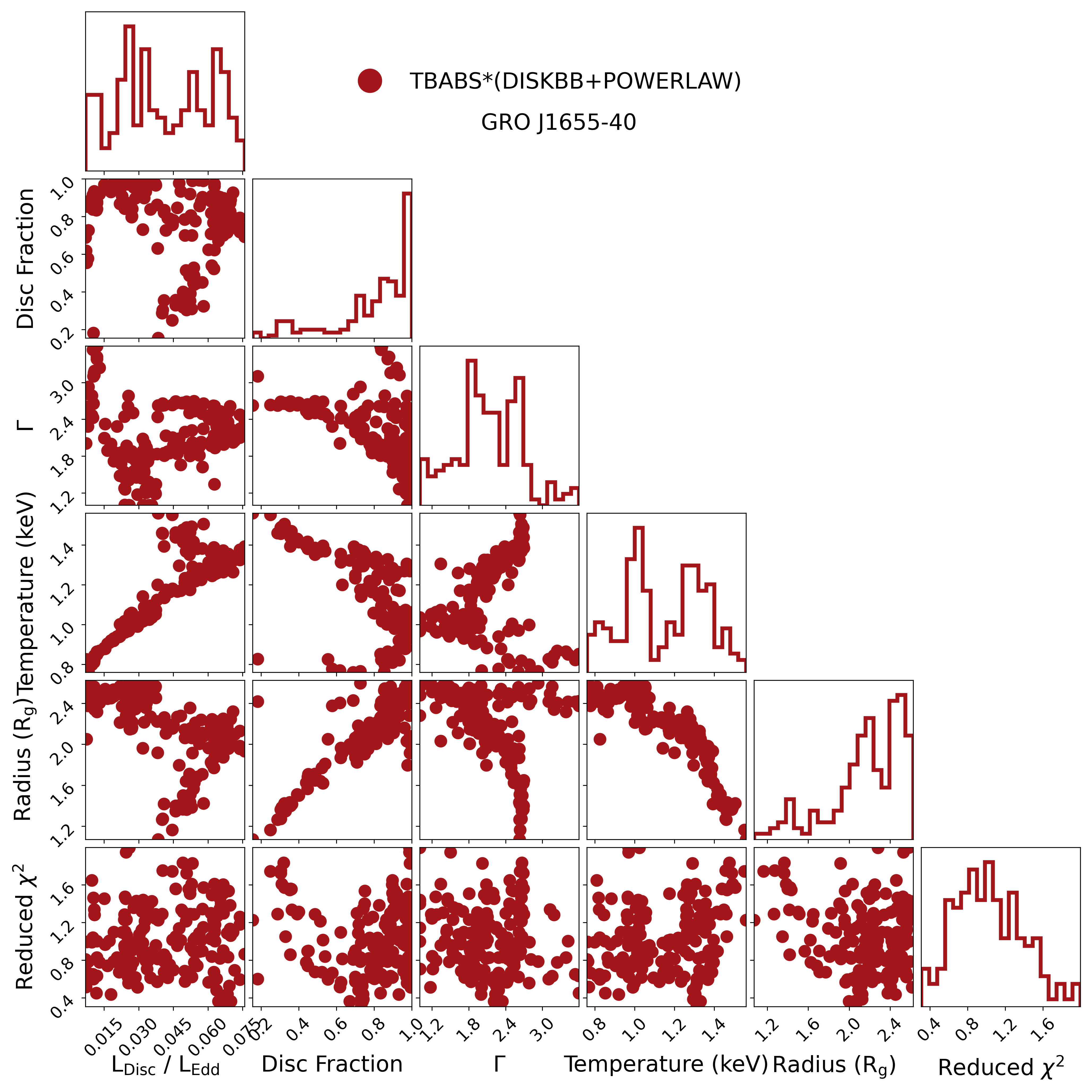}
        \subcaption{Corner plot showing the evolution of the parameters from our analysis of GRO J1655-40 with \texttt{TBABS$\times$(DISKBB+POWERLAW)}.}
        \label{fig:gro-all-diskbb}
    \end{subfigure}
    \caption{}
\end{figure*}
\begin{figure*}\ContinuedFloat
\medskip
    \begin{subfigure}{1.\textwidth}
    \centering
        \includegraphics[width=\textwidth]{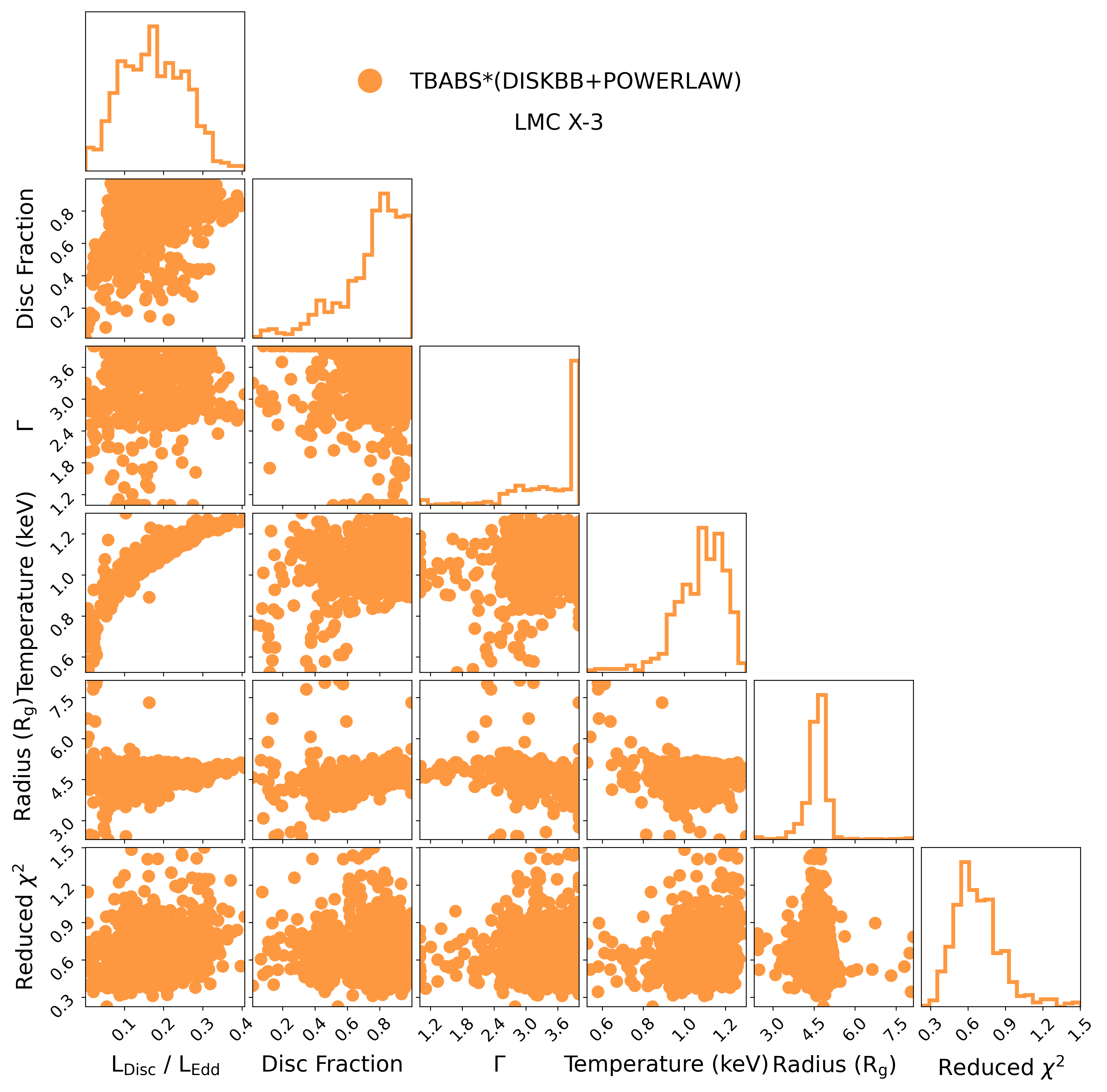}
        \subcaption{Corner plot showing the evolution of the parameters from our analysis of LMC X-3 with \texttt{TBABS$\times$(DISKBB+POWERLAW)}.}
        \label{fig:lmc-all-diskbb}
    \end{subfigure}
    \caption{(Continued)}
\end{figure*}
\begin{figure*}\ContinuedFloat
    \begin{subfigure}{1.\textwidth}
    \centering
        \includegraphics[width=\textwidth]{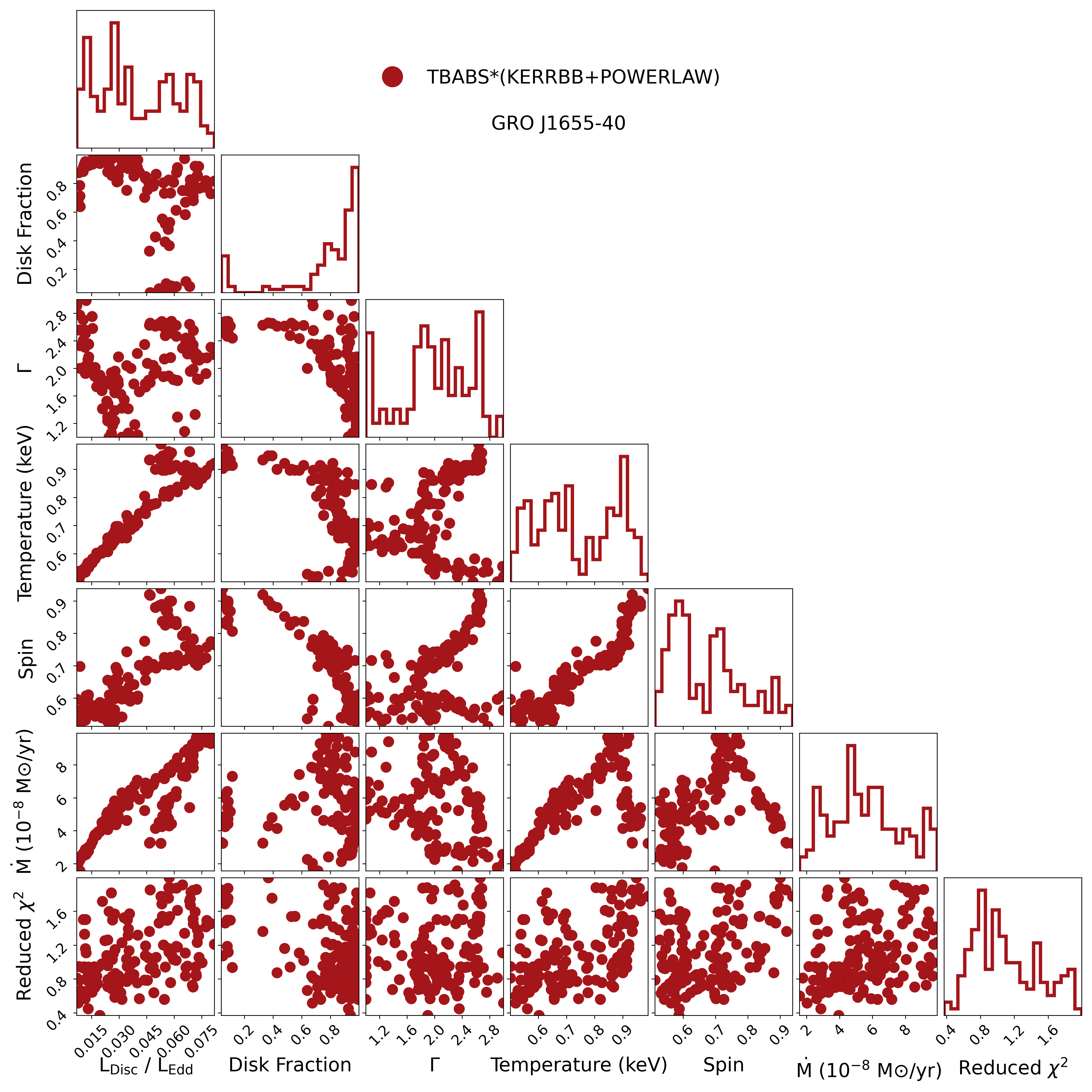}
        \subcaption{Corner plot showing the evolution of the parameters from our analysis of GRO J1655-40 with \texttt{TBABS$\times$(KERRBB+POWERLAW)}.}
        \label{fig:gro-all-kerrbb}
    \end{subfigure}
    \caption{(Continued)}
\end{figure*}
\begin{figure*}\ContinuedFloat
\medskip
    \begin{subfigure}{1.\textwidth}
    \centering
        \includegraphics[width=\textwidth]{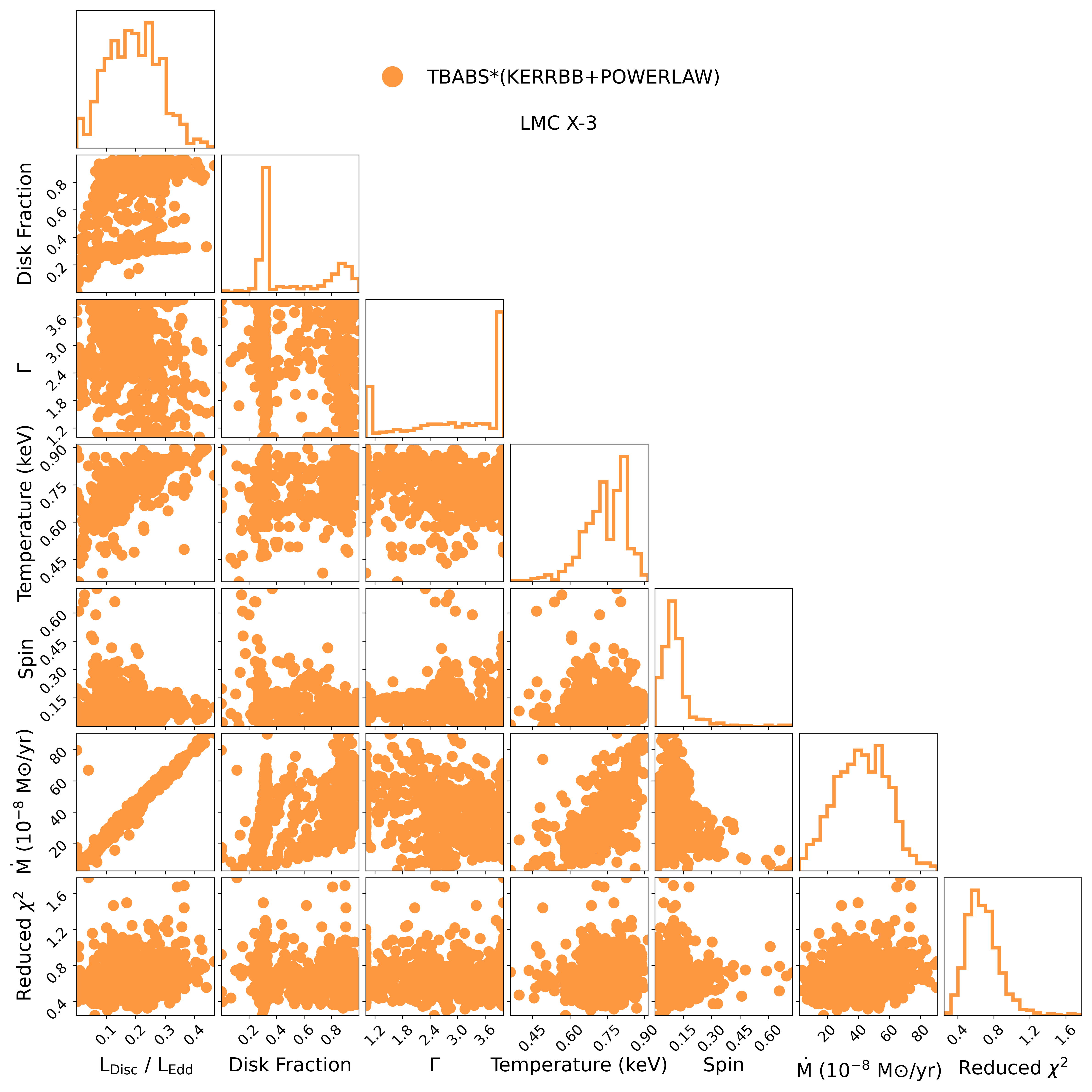}
        \subcaption{Corner plot showing the evolution of the parameters from our analysis of LMC X-3 with \texttt{TBABS$\times$(KERRBB+POWERLAW)}.}
        \label{fig:lmc-all-kerrbb}
    \end{subfigure}
    \caption{(Continued)}
\end{figure*}
\begin{figure*}\ContinuedFloat
    \begin{subfigure}{1.\textwidth}
    \centering
        \includegraphics[width=\textwidth]{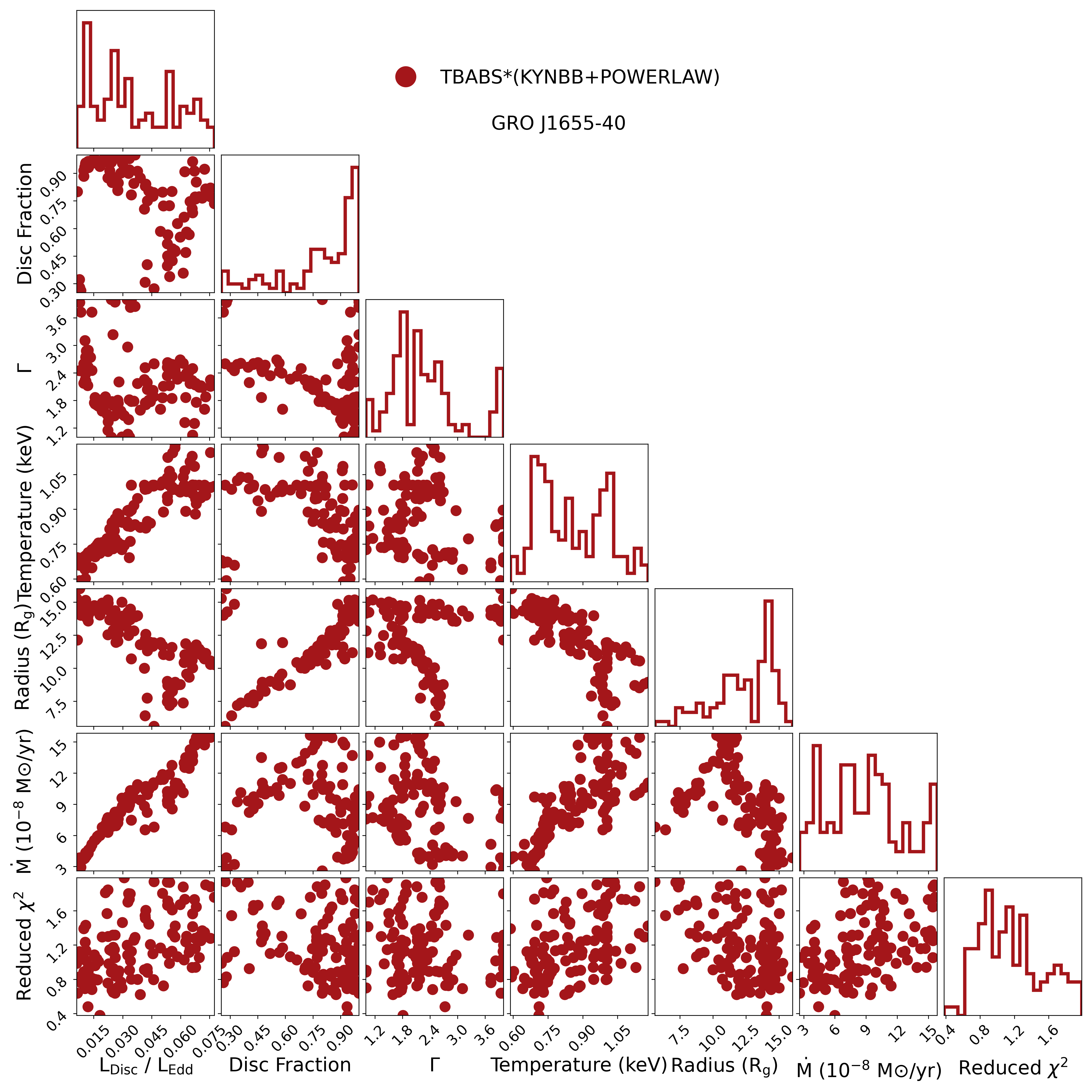}
        \subcaption{Corner plot showing the evolution of the parameters from our analysis of GRO J1655-40 with \texttt{TBABS$\times$(KYNBB+POWERLAW)}.}
        \label{fig:gro-all-kynbb}
    \end{subfigure}
    \caption{(Continued)}
\end{figure*}
\begin{figure*}\ContinuedFloat
\medskip
    \begin{subfigure}{1.\textwidth}
    \centering
        \includegraphics[width=\textwidth]{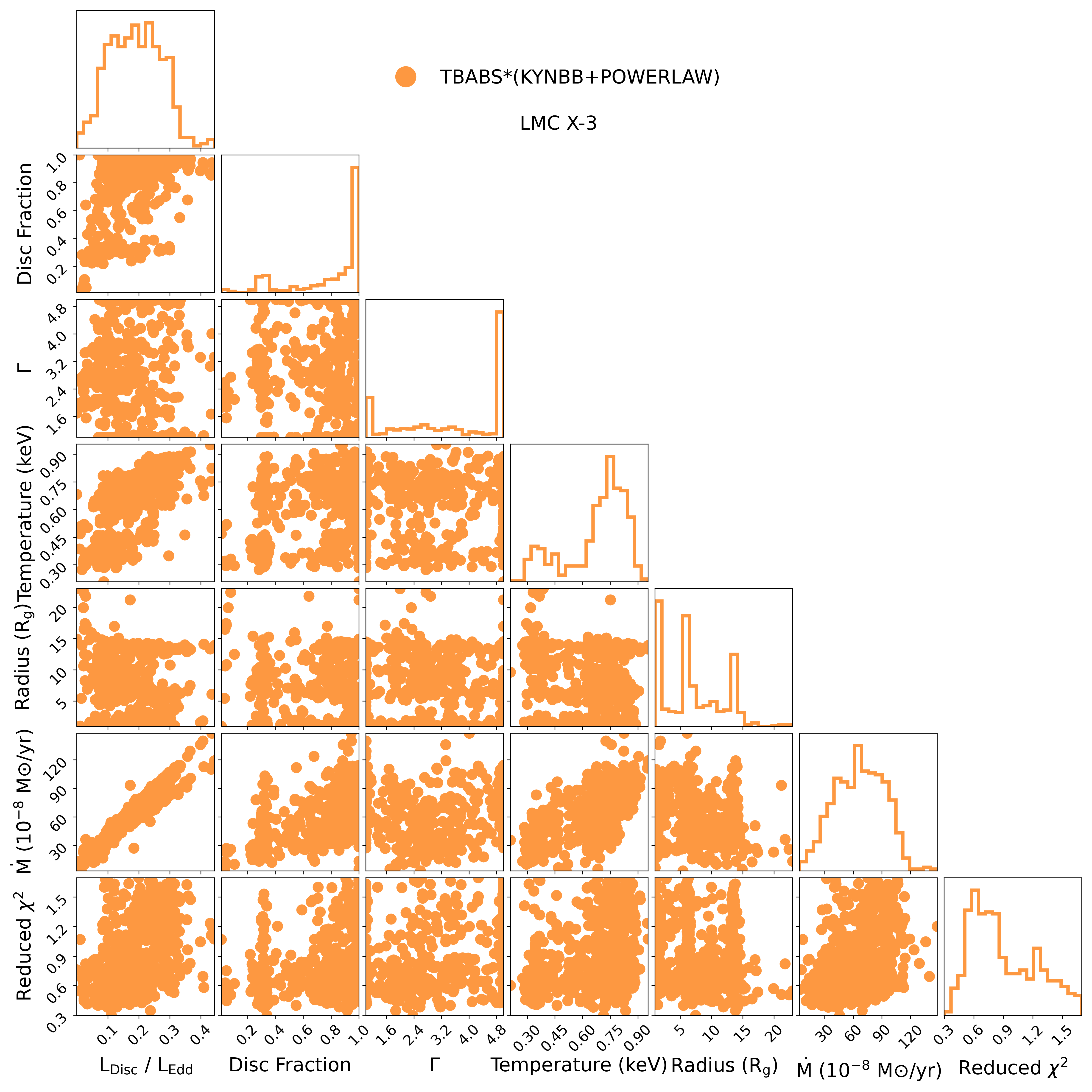}
        \subcaption{Corner plot showing the evolution of the parameters from our analysis of LMC X-3 with \texttt{TBABS$\times$(KYNBB+POWERLAW)}.}
        \label{fig:lmc-all-kynbb}
    \end{subfigure}
    \caption{(Continued)}
    \label{fig:all-both}
\end{figure*}


\bsp	
\label{lastpage}
\end{document}